\documentclass[journal=jctcce,manuscript=article,super=false]{achemso}
\usepackage[version=3]{mhchem} 
\usepackage{color}
\usepackage{multirow}
\usepackage{algorithm}
\usepackage{float}
\usepackage{longtable}
\usepackage{threeparttable}
\usepackage{ifthen}
\usepackage{array}
\usepackage{subfigure}
\usepackage{siunitx}
\usepackage{algpseudocode}
\usepackage{csquotes}
\usepackage{graphicx}
\usepackage{etoolbox}
\usepackage{pdfpages}
\usepackage[numbers]{natbib}

\makeatletter
\renewcommand\@cite[2]{[#1]}  
\makeatother

\newcommand{\onlinecite}[1]{Ref.~\citenum{#1}}

\author{Muhammad N. Tahir}
\affiliation{Institute of Physics, Chinese Academy of Sciences, Beijing 100190, China}
\email{bundesha@gmail.com}

\author{Honghui Shang}
       \affiliation{State Key Laboratory of Precision and Intelligent Chemistry, University of Science and Technology of China, Hefei 230026, China}
\email{shh@ustc.edu.cn}
      
\author{Xinguo Ren}
\email{renxg@iphy.ac.cn}
\affiliation{Institute of Physics, Chinese Academy of Sciences, Beijing 100190, China}
\alsoaffiliation{The NOMAD Laboratory at the Fritz Haber Institute of the Max Planck Society, Faradayweg 4-6, D-14195 Berlin, Germany}

\title[An \textsf{achemso} demo]
  {Analytical gradients of random-phase approximation plus corrections from renormalized single excitations}

\abbreviations{IR,NMR,UV}

\begin{document}
\newcommand{\kpe}{\mathbf{k}\!\cdot\!\mathbf{p}\,}
\newcommand{\Kpe}{\mathbf{K}\!\cdot\!\mathbf{p}\,}
\newcommand{\bfr}{ {\bf r}} 
\newcommand{\bfrp}{ {\bf r'}} 
\newcommand{\tp}{ {t^\prime}} 
\newcommand{\bfrpp}{ {\bf r^{\prime\prime}}} 
\newcommand{\bfR}{ {\bf R}} 
\newcommand{\bfq}{ {\bf q}} 
\newcommand{\bfp}{ {\bf p}} 
\newcommand{\bfk}{ {\bf k}} 
\newcommand{\bfG}{ {\bf G}} 
\newcommand{\bfGp}{ {\bf G'}} 
\newcommand{\dv}{ {\Delta \hat{v}}} 
\newcommand{\sigmap}{\sigma^\prime} 
\newcommand{\omegap}{\omega^\prime} 
\newcommand{\omegapp}{\omega^{\prime\prime}} 
\newcommand{\bracketm}[1]{\ensuremath{\langle #1   \rangle}}
\newcommand{\bracketw}[2]{\ensuremath{\langle #1 | #2  \rangle}}
\newcommand{\bracket}[3]{\ensuremath{\langle #1 | #2 | #3 \rangle}}
\newcommand{\ket}[1]{\ensuremath{| #1 \rangle}}
\newcommand{\GnWn}{\ensuremath{G_0W_0}\,}
\newcommand{\tn}[1]{\textnormal{#1}}
\newcommand{\f}[1]{\footnotemark[#1]}
\newcommand{\mc}[2]{\multicolumn{1}{#1}{#2}}
\newcommand{\mcs}[3]{\multicolumn{#1}{#2}{#3}}
\newcommand{\mcc}[1]{\multicolumn{1}{c}{#1}}
\newcommand{\refeq}[1]{(\ref{#1})} 
\newcommand{\refcite}[1]{Ref.~\cite{#1}} 
\newcommand{\refsec}[1]{Sec.~\ref{#1}} 
\newcommand\opd{d}
\newcommand\im{i}
\def\bra#1{\mathinner{\langle{#1}|}}
\def\ket#1{\mathinner{|{#1}\rangle}}
\newcommand{\braket}[2]{\langle #1|#2\rangle}
\def\Bra#1{\left<#1\right|}
\def\Ket#1{\left|#1\right>}

\newcommand{\fatr}{\mathbf{r}}

\newcommand{\XR}[1]{{\color{red}{\bf #1 }}}

\newcommand{\Or}{\mathcal{O}}
\newcommand{\ie}{\textit{i.e.}{}}
\renewcommand{\Im}{\mathrm{Im}~}
\newcommand{\Tr}{\mathrm{Tr}}

\newlength\replength
\newcommand\repfrac{.33}
\newcommand\dashfrac[1]{\renewcommand\repfrac{#1}}
\setlength\replength{8.5pt}
\newcommand\rulewidth{.6pt}

\begin{tocentry}\centering
 \includegraphics[width=0.5\textwidth]{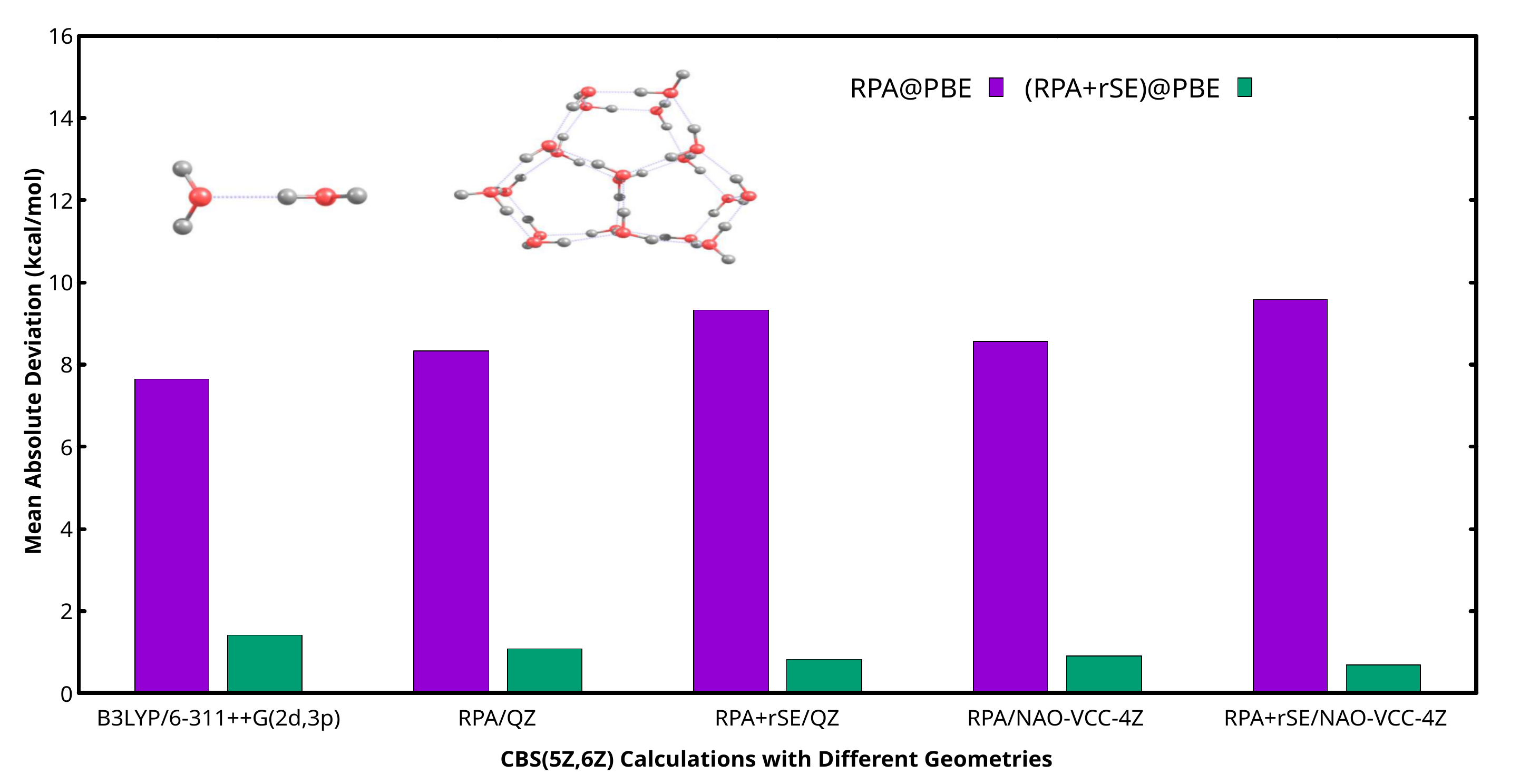}
\end{tocentry}

\begin{abstract}
 The random-phase approximation (RPA) formulated within the adiabatic connection fluctuation-dissipation framework is a powerful approach to compute the ground-state energies and properties of molecules and materials. Its overall underbinding behavior can be effectively mitigated by a simple correction term, called renormalized single excitation (rSE) correction. Analytical gradient calculations of the RPA energy have become increasingly available, enabling structural relaxations and even molecular dynamics at the RPA level. However, such calculations at the RPA+rSE level  have not been reported, due to the lack of the rSE analytical gradient. Here, we present the first formulation and implementation of the analytical gradients of the rSE energy with respect to the nuclear coordinates within an atomic-orbital basis set framework, which allows us to assess the performance of RPA+rSE in determining the molecular geometries and energetics. It is found that the slight overestimation behavior of RPA for small covalently bonded molecules is strengthened by rSE, while such behavior for molecules bonded with purely dispersion interactions is corrected. We further applied the approach to the water clusters, and found that the energy difference between the low-energy isomers of water hexamers is almost unchanged when going from RPA to RPA+rSE geometries. For the bigger WATER27 test set, using the RPA+rSE geometries instead of the RPA ones leads to a slight reduction of the mean absolute error of RPA+rSE from 0.91 kcal/mol to 0.70 kcal/mol, at the complete basis set.  
\end{abstract}

\section{Introduction}
Today, density functional theory (DFT) is the most popular method of choice for describing molecules and materials from first principles. However, the accuracy and reliability of DFT calculations depend crucially on the actual approximations to the underlying exchange-correlation (XC) functional. The XC approximations sitting on the first four rungs of the ``Jacob's ladder" \cite{Perdew/Schmidt:2001}, including local/semi-local \cite{Kohn/Sham:1965,Perdew/Burke/Ernzerhof:1996} and hybrid functionals \cite{Becke:1993,Heyd/Scuseria/Ernzerhof:2003}, do not capture nonlocal electron correlations which are indispensable for describing, e.g., van der Waals (vdW) interactions. 
Therefore, practical DFT calculations with conventional functionals have to be complemented with empirical or 
semi-empirical vdW corrections \cite{Grimme:2006b,Becke/Johnson:2007,Tkatchenko/Scheffler:2009} to describe systems where non-covalent interactions are important. However, for such schemes, a clean treatment of the intermediate, density-overlapping regime is always a challenge. 
A seamless, automatic inclusion of non-local electron correlations can be attained by going to the fifth rung of
the Jacob's ladder, where a prominent functional is the random phase approximation (RPA) \cite{Bohm/Pines:1953,Gell-Mann/Brueckner:1957} formulated via the framework of adiabatic connection fluctuation dissipation theorem (ACFDT) \cite{Langreth/Perdew:1977,Gunnarsson/Lundqvist:1976}. It has been shown that RPA provides a balanced description of different bonding interactions in complex chemical environments \cite{Ren/etal:2012b,Eshuis/Bates/Furche:2012}. In particular, RPA excels at discerning the ordering of the adsorption energies at different surface sites \cite{Ren/etal:2009,Schimka/etal:2010} and the energy differences between different isomers/polymorphs of the molecular and crystalline systems \cite{Lebegue/etal:2010,ZhangYubo/etal:2019,Sengupta/etal:2018,Zhang/Cui/Jiang:2018,Cazorla/Gould:2019,Yang/Ren:2022,Chedid/Jocelyn/Eshuis:2021,Muhammad_2022}.

Due to its attractive performance, it is natural to go beyond single-point energy calculations by enabling structural relaxation and molecular dynamics simulations based on the RPA potential energy surface. To this end, much effort has been devoted to developing computational formalisms and computer codes to evaluate the analytical gradient of RPA energies with respect to atomic displacements \cite{Rekkedal/etal:2013,Mussard/Szalay/Angyan/2014,Burow/Bates/Furche/Eshuis/2014,Ramberger/etal:2017,Beuerle/Ochsenfeld:2018,Chedid/Jocelyn/Eshuis:2021,Muhammad_2022,Muhammad_2024}. 
Efficient implementations of the RPA forces not only allow direct determination of the RPA 
structural and dynamical properties, but also facilitate the development of machine-learning force fields at the RPA level \cite{Liu/Verdi/etal:2012,Verdi/etal:2023}. 

However, the standard RPA scheme still suffers from some well-known shortcomings. Most notably, RPA shows a systematic underestimation of the binding energies of both molecules and solids. To address this issue, various schemes to correct the RPA have been developed, from various perspectives such as ACFDT \cite{Klimes/etal:2015,Erhard/etal:2016,Goerling:2019}, many-body perturbation theory (MBPT) \cite{Ren/etal:2011,Paier/etal:2012,Bates/Furche:2013}, time-dependent DFT \cite{Olsen/Thygesen:2012}, and
coupled cluster theory \cite{Grueneis/etal:2009,Paier/etal:2010}. Among these, a very simple and effective correction to RPA is the renormalized single excitation (rSE) correction scheme. The concrete expression of rSE can be derived from MBPT \cite{Ren/etal:2011,Ren/etal:2013} or the framework of ACFDT \cite{Klimes/etal:2015}, and it accounts for the variation of the density matrix along the adiabatic correction path. In terms of practical performance, RPA+rSE cures the underbinding problem of RPA to a large
extent for both molecules and solids \cite{Ren/etal:2012,Paier_2012,Ren/etal:2013}. It particularly shows excellent performance for describing the hydrogen 
bonding \cite{Ren/etal:2013}, on par with the coupled cluster method. It should be noted that rSE achieves this with a marginal extra cost. 
As such, RPA+rSE can be taken as a natural starting point to develop more sophisticated computational schemes \cite{Pham/etal:2024}.

In this connection, it would be highly desirable to calculate the analytical gradients at the level of RPA+rSE. To the best of our knowledge, such an implementation is not yet available.   Since the rSE term is computationally rather cheap,  one would expect that evaluating RPA+rSE gradients does not increase the computational overhead much. Efficient implementation of RPA+rSE gradients will allow us to examine whether the improved binding energies will lead to better structural properties. Previously, we have shown that the binding energies of water clusters have been significantly improved by adding the rSE correction to RPA \cite{Muhammad_2022,Muhammad_2024}. However, in these studies, the structures of the water clusters are still determined at the RPA level. The question is: What will happen if the structures are also consistently determined by RPA+rSE?

To address these questions, in this work, we develop the formalism and computer codes for calculating the 
rSE gradients with respect to the atomic displacements. Our implementation is built upon the previously developed formalism for the RPA gradients \cite{Muhammad_2022,Muhammad_2024} within the framework of atomic orbital (AO) basis sets. Adding these two components together, we can have access to the RPA+rSE forces, allowing us to relax the molecular structures at the RPA+rSE level.

This paper is structured as follows: The theoretical formulation is presented in Sec.~\ref{sec:theory} where we begin by briefly reviewing the energy formula of RPA and rSE, and this is followed by a detailed derivation of the formulation of the analytical gradient of the rSE correlation energy with respect to atomic displacements. The implementation and computational details are given in Sec.~\ref{sec:imple}. In Sec.~\ref{sec:results}, we present the main results of this work, including the benchmark tests against the finite difference results, the structural properties of small molecules, and the effect of structures on the energetics for the water clusters.  Finally, we conclude this work in Sec.~\ref{sec:conclusion}.

\section{\label{sec:theory}Theoretical Formulation}

In this section, we first present the key equations behind the RPA+rSE computational method \cite{Ren/etal:2013}, based on the resolution of identity (RI) formalism \cite{Ihrig/etal:2015,Levchenko/etal:2015,Lin/Ren/He:2020}, which has been implemented in the FHI-aims \cite{Blum/etal:2009,Havu/etal:2009,Ren/etal:2012} code. This is followed by an exposition of the formalism of the gradients of the RPA+rSE total energy with respect to the atomic displacements, whereby the new terms that need to be evaluated will be highlighted.

\subsection{The RPA+rSE total energy}
The RPA+rSE total energy is given by
\begin{equation}
 E^{\text{RPA+rSE}}=E^{\text{DFA}}-E^{\text{DFA}}_{xc}+E^{\text{EX}}_{x}+E^{\text{RPA}}_{c}+E^{\text{rSE}}_{c}\, 
 \label{eq:RPArSE_total_energy}
\end{equation}
where $E^{\text{DFA}}$ is the total energy under  conventional density functional approximation (DFA), usually
at the level of local-density or generalized gradient approximations (LDA/GGAs), and $E^\text{DFA}_{xc}$ is the corresponding exchange-correlation (xc) energy. Furthermore, $E^{\text{EX}}_{x}$,  $E^{\text{RPA}}_{c}$, and
$E^{\text{rSE}}_{c}$ are, respectively, the exact-exchange (EX) energy, the RPA correlation energy, and
the rSE correlation energy. Similar to the standard RPA case, in usual RPA+rSE calculations, one first performs a DFA
calculation, and then evaluates the EX energy and RPA+rSE correlation energy in a post-processing manner using single-particle orbitals and orbital energies generated from the preceding DFA calculation.  For completeness, here
we will briefly recapitulate the expressions of the key terms to set up the stage. In the following, we use
 $m,n,o$ to label occupied molecular orbitals (MOs), $a,b,u$ for unoccupied MOs, and $p,q,r,s$ for general ones. Furthermore, $i,j,k,l$ are used for atomic orbital (AO) basis and $\mu,\nu$ for auxiliary basis functions (ABFs). For clarity, in this work we often choose to explicitly specify the summation range of the MOs.
 The EX energy is given by
\begin{equation}
      E^{\text{EX}}_{x}=-\sum_{m,n}^{occ} (mn|nm)\, ,
      \label{eq:E_EX}
\end{equation}
where $(mn|nm)$ is the two-electron Coulomb repulsion integral (ERIs) in terms of the KS MOs. These ERIs 
are in general defined as
\begin{equation}
    (pq|rs)=\int \frac{\psi_p(\bf{r})\psi_q(\bf{r})\psi_r(\bf{r'})\psi_s(\bf{r'})}{|\bf{r}-\bf{r'}|} d\bf{r} d\bf{r'} \, .
    \label{eq:ERI}
\end{equation}
In Eq.~\ref{eq:RPArSE_total_energy}, the RPA correlation energy can be conveniently calculated as
\begin{equation}
    E_c^\text{RPA} = \frac{1}{2\pi} \int_0^\infty d\omega \text{Tr}\left[\text{ln}\left(\textbf{1}-\boldsymbol{\chi}^0 (i\omega) \boldsymbol{v}\right) + \boldsymbol{\chi}^0(i\omega) \boldsymbol{v} \right]\,,
\end{equation}
where $\boldsymbol{\chi}^0$ and $\boldsymbol{v}$ should be understood as the matrix form of the noninteracting
density response function and the bare Coulomb interaction, represented in terms of  a set of ABFs. In real space,
$\boldsymbol{\chi}^0$ is given by,

\begin{equation}
\boldsymbol{\chi}^0(\textbf{r},\textbf{r}^{\prime},i\omega) = \sum_{p,q} \frac{(\eta_p-\eta_q)\psi_p(\bf{r})\psi_q(\bf{r})\psi_q(\bf{r'})\psi_p(\bf{r'})}{\epsilon_p - \epsilon_q - i\omega}\, ,
\label{eq:chi0_realspace}
\end{equation}

where $\psi_p$, $\epsilon_p$, and $\eta_p$ are the KS MOs, their energies, and occupation numbers, respectively. For simplicity, here we assume closed-shell systems and real orbitals, but extending to
more general cases is straightforward.

Within the AO basis set framework, the KS MOs are expanded in terms of the AOs $\{\phi_i(\bf{r})\}$,
\begin{equation}
    \psi_p(\textbf{r}) = \sum_i c_{i,p} \phi_i(\bf{r})
\end{equation}
with $c_{i,p}$ being the KS eigenvectors.
In practical implementations, the RI approximation is commonly used in the computation of the EX and RPA correlation energies. Within this approximation, 
a set of ABFs $\{P_\mu(\bf{r})\}$ are introduced to expand the products of two AOs, 
\begin{equation}
    \phi_i(\textbf{r})\phi_j(\textbf{r})=\sum_{\mu}C_{ij}^\mu P_\mu(\textbf{r})
    \label{eq:RI_AO_expansion}
\end{equation}
where $C_{ij}^\mu$ are the expansion coefficients. Moreover, the Coulomb operator is also represented
in the basis of the ABFs, yielding the so-called Coulomb matrix
\begin{equation}
    V_{\mu\nu} = \int d\textbf{r} d\textbf{r}^{\prime} \frac{P_\mu(\textbf{r})P_\nu(\textbf{r}^
    {\prime})}{|\textbf{r}-\textbf{r}^{\prime}|}\, .
\end{equation}

Introducing an intermediate quantity 
\begin{equation}
    O_{pq}^\mu = \sum_{ij}c_{i,p}Q_{ij}^\mu c_{j,q}\, ,
    \label{eq:O_integrals}
\end{equation}
with
\begin{equation}
    Q_{ij}^\mu =\sum_\nu C_{ij}^\nu \left[V^{\frac{1}{2}}\right]_{\nu,\mu}\, ,
    \label{eq:Q_integrals}
\end{equation}
we can obtain the computational expressions for the EX energy 
\begin{equation}
E_{x}^{\text{EX}}=-\sum_{mn}^{occ} \sum_{\mu}O^{\mu}_{mn} O^{\mu}_{nm} \, ,
\label{eq:EX_RI}
\end{equation}
and for RPA correlation energy 
\begin{equation} 
E_c^\text{RPA} = \frac{1}{2\pi} \int_0^\infty d\omega \text{Tr}\left[\text{ln}\left(\textbf{1}-\mathbf{\Pi}(i\omega)\right) + \mathbf{\Pi}(i\omega) \right]\, ,
\label{eq:EcRPA_RI}
\end{equation}
under the RI approximation, respectively.
In Eq.~\ref{eq:EcRPA_RI},

\begin{equation}
   \mathbf{\Pi}_{\mu\nu} (i\omega)=\sum_{p,q}\frac{(\eta_p-\eta_q)O_{pq}^\mu O_{qp}^\nu}{\epsilon_p-\epsilon_q-i\omega}\, .
   \label{eq:Pi_matrix}
\end{equation}
More detailed derivations of the above equations, on which our RPA force implementation is based, can be found in
Refs.~\cite{Ren/etal:2012,Ren/etal:2012b}. \\

The final piece in the RPA+rSE total energy (Eq.~\ref{eq:RPArSE_total_energy}) is the rSE correlation energy, which is formally given by

\begin{equation}
    E_c^\text{rSE} = \sum_{o}^\text{occ}\sum_u^\text{unocc} \frac{|F_{ou}|^2}{\lambda_o - \lambda_u}
    \label{eq:rSE_energy}
\end{equation}
where the energies $\lambda_o$ and $\lambda_u$ in the denominator of Eq.~\ref{eq:rSE_energy} are obtained by solving the eigenvalue problems 
\begin{equation}
    \sum_{n}^{occ}f_{mn}{\cal O}_{no} = {\cal O}_{mo}\lambda_o
    \label{eq:eigen_occ}
\end{equation}
and 
\begin{equation}
       \sum_{b}^{unocc}f_{ab}{\cal U}_{bu} = {\cal U}_{au}\lambda_u \, .
      \label{eq:eigen_unocc}
\end{equation}
In the above equations, $f_{mn}$ and $f_{ab}$ are the occupied and unoccupied blocks of the Fock matrix (within KS orbital basis), and  $\{\lambda_o, {\cal O}_{mo}\}$ and  $\{\lambda_u, {\cal U}_{au}\}$ are
the eigen-pairs of the two blocks, respectively. This procedure is usually referred to as semi-canonicalization and and the resultant orbitals (defined by the eigenvectors ${\cal O}_{mo}$ and ${\cal U}_{au}$)
are called semi-canonical orbitals.
The numerator $F_{ou}$ in Eq.~\ref{eq:rSE_energy} is given by
\begin{equation}
    F_{ou} = \sum_{m}^{occ}\sum_{a}^{unocc} {\cal O}_{mo}^\ast f_{ma} {\cal U}_{au}\, 
    \label{eq:Fou}
\end{equation}
which can be seen as the occupied-unoccupied block of the Fock matrix represented in the semi-canonical orbital basis.

Using the KS orbitals, we define the Fock operator $f$ as 
\begin{equation}
    \hat{f}=-\frac{\nabla^2}{2} + \hat{v}_\text{ext} + \hat{v}_\text{H} + \hat{v}_\text{x}^\text{HF}= \hat{h}_\text{KS} -\hat{v}_\text{xc} +\hat{v}_\text{x}^\text{HF} 
\end{equation}
whereby the KS Hamiltonian operator
\begin{equation}
\hat{h}_\text{KS}=-\frac{\nabla^2}{2} + \hat{v}_\text{ext} + \hat{v}_\text{H} + \hat{v}_\text{xc}\, .
\end{equation}

The Fock matrix in the basis of KS MOs is defined as  
\begin{align}
    f_{pq}&= \langle \psi_{p} |\hat{f} |\psi_{q} \rangle = \langle \psi_{p} |\hat{v}^\text{HF}_\text{x} - \hat{v}_\text{xc} +\hat{h}_\text{KS} |\psi_{q} \rangle \nonumber \\
    & =-\sum_{m}^{occ} \langle pm  | mq\rangle - V_{pq}^\text{xc}+ \epsilon_{p} \delta_{pq}\, .
    \label{eq:fock_matrix}
\end{align}
where $\langle \psi_{p} |\hat{h}_\text{KS} |\psi_{q}\rangle = \epsilon_{p}\delta_{p,q}$ is used. 
In Eq.~\ref{eq:fock_matrix}, $\langle pm|mq\rangle$ is the ERIs defined via Eq.~\ref{eq:ERI}

and
\begin{equation}
V_{pq}^\text{xc} = \sum_{i,j}c_{i,p}^\ast V_{ij}^\text{xc}c_{j,q}
\label{eq:VxcAO2MO}
\end{equation}
with
\begin{equation}
    V_{ij}^\text{xc}=\int \phi_{i}(\textbf{r}) \boldsymbol{v}_\text{xc}(\textbf{r}) \phi_{j}(\textbf{r})d\textbf{r} \, .
\end{equation}

Finally, within the RI approximation, we have 
\begin{equation}
    f_{pq}=-\sum_{\mu}\sum_{m}^{occ} O_{pm}^\mu O_{mq}^\mu -V^\text{xc}_{pq}+\delta_{pq}\epsilon_{p} \, .
    \label{eq:f_matr_in_KS}
\end{equation}

\subsection{The RPA + rSE analytical forces}

The RPA+rSE force is given by the gradient of the RPA+rSE total energy with respect to the atomic displacement,
 \begin{equation} \label{eq:RPA_rSE_force_total}
 \begin{split}
 \mathbf{F}_{A}^{\text{RPA+rSE}}&=-\dfrac{dE^\text{RPA+rSE}}{d\mathbf{R}_{A}}\\&
 =-\dfrac{dE^{\text{DFA}}}{d\mathbf{R}_{A}}+\dfrac{dE^{\text{DFA}}_{xc}}{d\mathbf{R}_{A}}-\dfrac{dE^{\text{EX}}_{x}}{d\mathbf{R}_{A}} \\&
~~ ~~-\dfrac{dE^{\text{RPA}}_{c}}{d\mathbf{R}_{A}}-\dfrac{dE^{\text{rSE}}_{c}}{d\mathbf{R}_{A}}\, 
 \end{split}
 \end{equation}
 where $\bf{R}_A$ denotes the nuclear position of a given atom $A$.
In Eq.~\ref{eq:RPA_rSE_force_total}, the first term -- the force at the level of conventional DFAs has long been available in FHI-aims. The remaining four terms are those that need to be evaluated in our RPA+rSE force implementation. 

In the following, we will mainly discuss the rSE part of
the total force, since the computational framework and computer code for the other three terms have been
recently developed \cite{Muhammad_2022,Muhammad_2024} and are currently available in FHI-aims \cite{Blum/etal:2009}.

Nevertheless, from a computational point of view, it is often convenient to group the EX  and correlation contributions
together, and for this purpose, we introduce a keyword (P-XC) for the total force arising from 
the EX energy and RPA+rSE correlation energy, 
\begin{equation}
\begin{split}
    \mathbf{F}^{\text{P-XC}}_{A}&= \mathbf{F}^{\text{EX}}_{x,A}+\mathbf{F}^{\text{RPA}}_{c,A}+\mathbf{F}^{\text{rSE}}_{c,A}\\&
    =-\dfrac{d~~~}{d\mathbf{R}_{A}} \Big(E^{\text{EX}}_{x} +E^{\text{RPA}}_{c} +E^{\text{rSE}}_{c}\Big) \, .
    \end{split}
\end{equation}

The EX, RPA, and rSE energies directly depend on the KS eigenvectors ($c$), Coulomb matrix $V$, and the RI expansion coefficients ($C$). For RPA and rSE, there is an additional dependence on the KS eigenvalues $\epsilon$, and for the rSE alone, there is an additional dependence on the KS XC potential \( V_{xc} \) (see Eq.~\ref{eq:f_matr_in_KS}). Naturally, by applying the chain rule in the trace operation, the force can also be split into five parts,
\begin{equation}
\begin{split}
\mathbf{F}^{\text{P-XC}}_{A}=&-\big\langle \Gamma^{(1)}_\text{P-XC} \dfrac{dc}{d\mathbf{R}_{A}} \big \rangle   -\big\langle \Gamma^{(2)}_\text{P-XC} \dfrac{d\epsilon}{d\mathbf{R}_{A}} \big \rangle  \\& 
- \big\langle \Gamma^{(3)}_\text{P-XC} \dfrac{dC}{d\mathbf{R}_{A}}\big \rangle - \big\langle \Gamma^{(4)}_\text{P-XC} \dfrac{dV}{d\mathbf{R}_{A}} \big \rangle \\& 
-\big\langle \Gamma^{(5)}_\text{P-XC} \dfrac{dV^{xc}}{d\mathbf{R}_{A}} \big \rangle\, ,
\end{split}
\label{eq:force_PXC_five_terms}
\end{equation}

where $\left\langle \right\rangle$ stands for the trace operation over the intermediate variables. Now, the $\Gamma_\text{P-XC}$ matrix contains three components: $\Gamma_{EX-x}$, $\Gamma_{RPA-c}$, and $\Gamma_{rSE-c}$,
to which not all of the five terms in Eq.~\ref{eq:force_PXC_five_terms} have non-zero contributions. 
For instance, $\Gamma^{(2)}_{EX-x}=0$ because the EX  energy does not involve KS eigenvalues ($\epsilon$). Similarly, $\Gamma^{(5)}=0$ for the EX and RPA correlation energies since they don't have an explicit dependence on DFA XC potential.

As alluded to above, our emphasis here is primarily on $\Gamma^\text{rSE}_c$.
Formally, the analytical gradient of the rSE correlation energy with respect to the atomic positions is given by
 \begin{equation}
 \begin{split}
     \mathbf{F}^{\text{rSE}}_{A}&= -\dfrac{dE^{\text{rSE}}_{c}}{d\textbf{R}_{A}} \\& 
     = - \langle \dfrac{dE^{\text{rSE}}_{c}}{dc} \dfrac{dc}{d\textbf{R}_{A}}\rangle 
  -\langle \dfrac{dE_{c}^{\text{rSE}}}{d\epsilon} \dfrac{d\epsilon}{d\textbf{R}_{A}}\rangle 
 - \langle \dfrac{dE^{\text{rSE}}_{c}}{dC} \dfrac{dC}{d\textbf{R}_{A}}\rangle  \\ & 
~~~~    -\langle \dfrac{dE^{\text{rSE}}_{c}}{dV} \dfrac{dV}{d\textbf{R}_{A}}\rangle  -
     \langle \dfrac{dE^{\text{rSE}}_{c}}{dV^\text{xc}} \dfrac{dV^\text{xc}}{d\textbf{R}_{A}}\rangle \\& 
     = - \sum_{\zeta= \{c,\epsilon,C,V,V^\text{xc}\}} \langle \dfrac{dE^{\text{rSE}}_{c}}{d\zeta} \dfrac{d\zeta}{d\textbf{R}_{A}}\rangle \\& 
     = - \langle\Gamma^{(1)} \dfrac{dc}{d\textbf{R}_{A}}\rangle - \langle\Gamma^{(2)} \dfrac{d\epsilon}{d\textbf{R}_{A}}\rangle  
- \langle \Gamma^{(3)} \dfrac{dC}{d\textbf{R}_{A}}\rangle \\&
  ~~~~ -  \langle \Gamma^{(4)}\dfrac{dV}{d\textbf{R}_{A}}\rangle 
    - \langle \Gamma^{(5)} \dfrac{dV^\text{xc}}{d\textbf{R}_{A}}\rangle
  \end{split}
  \label{eq:rse_gradient}
 \end{equation}

Here we use $\zeta$ to denote one of the five intermediate variables, $c$, $V$, $C$, $V^\text{xc}$, and $\epsilon$. These are either vectors ($\epsilon$) or rank-2 ($c,V,V^\text{xc}$) and rank-3 tensors ($C$). The derivatives in
Eq.~\ref{eq:rse_gradient} are taken with respect to each element in these vectors/tensors, and the trace is taken over all their indices.
The derivative of the rSE energy with respect to an unspecified variable $\zeta$,
 despite being highly complex, is formally rather general, which we shall discuss first. Using Eqs.~\ref{eq:rSE_energy} and \ref{eq:Fou}, and following the chain rules, we have
\begin{equation} \label{eq:dc_force}
\begin{split} 
  \dfrac{dE^{rSE}_{c}}{d\zeta} &= \dfrac{d}{d\zeta} \Bigg[ \sum_{o}^{occ} \sum_{u}^{unocc} \dfrac{|F_{ou}|^{2}}{\lambda_{o}-\lambda_{u}}\Bigg] \\& 
 =2\sum_{o}^{occ}\sum_{u}^{unocc} \dfrac{F_{ou}}{\lambda_{o}-\lambda_{u}} \dfrac{dF_{ou}}{d\zeta} +\sum_{o}^{occ}\sum_{u}^{unocc} \dfrac{|F_{ou}|^{2}}{(\lambda_{o}-\lambda_{u})^{2}} \big ( \dfrac{d\lambda_{u}}{d\zeta}-\dfrac{d\lambda_{o}}{d\zeta}\big) \\& 
   = 2\sum_{o,m}^{occ}\sum_{u,a}^{unocc} \dfrac{F_{ou}f_{ma}}{\lambda_{o}-\lambda_{u}} \Big[\dfrac{d{\cal O}^{\star}_{mo}}{d\zeta} {\cal U}_{au} +{\cal O}^{\star}_{mo}\dfrac{d{\cal U}_{au}}{d\zeta} \Big] \\& 
  +2\sum_{o,m}^{occ}\sum_{u,a}^{unocc} \dfrac{{\cal O}^{\star}_{mo} F_{ou}{\cal U}_{au}}{\lambda_{o}-\lambda_{u}}  \dfrac{df_{ma}}{d\zeta}  +  \sum_{o}^{occ}\sum_{u}^{unocc} \dfrac{|F_{ou}|^{2}}{(\lambda_{o}-\lambda_{u})^{2}} \big ( \dfrac{d\lambda_{u}}{d\zeta}-\dfrac{d\lambda_{o}}{d\zeta}\big) \, .  
  \end{split}
\end{equation}
Now, the determination of the rSE force component boils down to the computation of  the first-order derivatives of $f$, $\lambda$,  ${\cal O}$, and ${\cal U}$ with respect to the variable $\zeta$.
 For brevity, we  denote $\dfrac{df}{d\zeta}=f^{(1)}_{\zeta}=f^{(1)}$, $\dfrac{d\lambda}{d\zeta}=\lambda^{(1)}$, $\dfrac{d{\cal O}}{d\zeta}={\cal O}^{(1)}$ and $\dfrac{d{\cal U}}{d\zeta}={\cal U}^{(1)}$, and omit the superscript `rSE' without causing confusion. 
 To determine these derivatives, we differentiate both sides of Eq.~(\ref{eq:eigen_occ}) with respect to the 
 $\zeta$, and this gives rise to
\begin{equation}
    \sum_{n}\left(f_{mn} - \delta_{mn} \lambda_o \right){\cal O}^{(1)}_{no} = \sum_{n}\left(\lambda_o^{(1)}\delta_{mn} -  f_{mn}^{(1)}\right){\cal O}_{no}
    \label{eq:dfpt_occ_1}
\end{equation}
To solve Eq.~(\ref{eq:dfpt_occ_1}), it is convenient to expand ${\cal O}^{(1)}_{no}$ in terms of the corresponding zeroth-order eigenvectors, i.e.,
\begin{equation}
    {\cal O}^{(1)}_{mo} =\sum_{o'}^{occ} {\cal O}_{mo'}X^{(1)}_{o'o} \,  
\end{equation}
and a simple derivation yields 
\begin{equation}
    X^{(1)}_{o'o} =-\dfrac{\sum_{mn}^{occ} {\cal O}_{mo'}^\ast f_{mn}^{(1)} {\cal O}_{no}}{\lambda_{o'} - \lambda_o}\,,  \quad\quad o\ne o' \, .
\end{equation}
Furthermore, it is straightforward to show that
\begin{equation}
    \lambda_o^{(1)} = \sum_{mn}^{occ} {\cal O}_{mo}^\ast f_{mn}^{(1)}{\cal O}_{no}
\end{equation}

Similarly, for the unoccupied block, one has 
\begin{equation}
    {\cal U}^{(1)}_{au} =\sum_{u'}^{unocc} {\cal U}_{au'}X^{(1)}_{u'u}\, ,
\end{equation}
with
\begin{equation}
    X^{(1)}_{u'u} =-\dfrac{\sum_{ab}^{unocc} {\cal U}_{au'}^\ast f_{ab}^{(1)} {\cal U}_{bu}}{\lambda_{u'} - \lambda_u}\,, \quad\quad u\ne u' \, ,
\end{equation}
 and
\begin{equation}
    \lambda_u^{(1)} = \sum_{ab}^{unocc} {\cal U}_{au}^\ast f_{ab}^{(1)}{\cal U}_{bu}\, .
\end{equation}

The above equations show that the first-order quantities  $\lambda^{(1)}$, ${\cal O}^{(1)}$ and ${\cal U}^{(1)}$ can all be expressed in terms of their zeroth order counterparts and the first-order Fock matrix $f^{(1)}$. 
Combining these equations, we can rewrite Eq.~(\ref{eq:dc_force}) as 
\begin{equation}
\begin{split}
  \dfrac{dE^{rSE}_{c}}{d\zeta}=& -2\sum_{o,m}^{occ}\sum_{u,a}^{unocc} \dfrac{F_{ou}f_{ma}}{\lambda_{o}-\lambda_{u}}  
  \Big[\sum_{m',n,o'}^{occ} \dfrac{{\cal O}_{mo'}  {\cal U}_{au}{\cal O}_{m'o'}^\ast f_{m'n,\zeta}^{(1)} O_{no}}{\lambda_{o'} - \lambda_o}  +\sum_{a',b,u'}^{unocc}  \dfrac{{\cal O}^{\star}_{mo} {\cal U}_{a'u'}^\ast f_{a'b,\zeta}^{(1)} {\cal U}_{bu} {\cal U}_{au'}}{\lambda_{u'} - \lambda_u} \Big] \\&
  +\sum_{o}^{occ}\sum_{u}^{unocc} \dfrac{|F_{ou}|^{2}}{(\lambda_{o}-\lambda_{u})^{2}} 
  \Big [ \sum_{a,b}^{unocc} {\cal U}_{au}^\ast f_{ab, \zeta}^{(1)}{\cal U}_{bu} 
  -\sum_{m,n}^{occ} {\cal O}_{mo}^\ast f_{mn, \zeta}^{(1)}{\cal O}_{no}\Big] \\& 
   +2\sum_{o,m}^{occ}\sum_{u,a}^{unocc} \dfrac{{\cal O}^{\star}_{mo} F_{ou}{\cal U}_{au}}{\lambda_{o}-\lambda_{u}}  f^{(1)}_{ma, \zeta}\, ,
  \end{split}
  \label{eq:dgammabydx}
\end{equation}
where the dependence of $f^{(1)}$ matrices on $\zeta$ is explicitly indicated.

In compact form, Eq.~\ref{eq:dgammabydx} can be re-expressed as
\begin{equation} 
\begin{split}
  \dfrac{dE^{rSE}_{c}}{d\zeta} &= \sum_{m,n}^{occ} {\cal F}_{mn}  f_{mn, \zeta}^{(1)}  +\sum_{a,b}^{unocc}  {\cal F}_{ab} f_{ab,\zeta}^{(1)} +\sum_{m}^{occ}\sum_{a}^{unocc} {\cal F}_{ma} f^{(1)}_{ma, \zeta} \\
  &= \sum_{p,q} \Lambda_{pq}\dfrac{f_{pq}}{d\zeta}\, ,
  \end{split}
  \label{eq:generalized_formula} 
\end{equation}
with  
\begin{equation} \label{eq:occ_occ_block}
\begin{split}
    {\cal F}_{mn}=&  -2\sum_{m',n',o}^{occ} \Bigg( \sum_{a,u}^{unocc}\dfrac{f_{m'a} F_{ou} {\cal U}_{au}}{\lambda_{o}-\lambda_{u}}\Bigg)
     {\cal O}_{m'n'}\dfrac{{\cal O}_{mn'} {\cal O}_{no}}{\lambda_{n'}-\lambda_{o}} \\& 
     -\sum_{o}^{occ} \sum_{u}^{unocc}  {\cal O}_{mo}\dfrac{|F_{ou}|^{2}}{(\lambda_{o}-\lambda_{u})^{2}} {\cal O}_{no} 
     \end{split}
\end{equation}
for the occupied-occupied block,
\begin{equation} \label{eq:unocc_unocc_block}
\begin{split}
   {\cal F}_{ab} =& -2\sum_{a',b',u}^{unocc} \Bigg(\sum_{m,o}^{occ}\dfrac{{\cal O}_{mo}F_{ou}f_{ma'}  }{\lambda_{o}-\lambda_{u}}\Bigg) 
  {\cal U}_{a'b'}\dfrac{{\cal U}_{ab'} {\cal U}_{bu}}{\lambda_{b'}-\lambda_{u}} \\& 
  +\sum_{o}^{occ} \sum_{u}^{unocc} {\cal U}_{au}\dfrac{|F_{ou}|^{2}}{(\lambda_{o}-\lambda_{u})^{2}} {\cal U}_{bu}\,
  \end{split}
\end{equation}
for the unoccupied-unoccupied block, and
\begin{equation} \label{eq:occ_unocc_block}
    {\cal F}_{ma}= 2\sum_{o}^{occ} \sum_{u}^{unocc}\dfrac{{\cal O}_{mo} F_{ou} {\cal U}_{au}}{\lambda_{o}-\lambda_{u}} \,
\end{equation}
for the occupied-unoccupied block.
In Eq.~\ref{eq:generalized_formula}, for convenience, we have introduced a supermatrix 
\begin{equation}
 { \bf{ \Lambda}} =
   \begin{pmatrix}
 {\cal F}_{mn} &  {\cal F}_{ma} \\
 0 &  {\cal F}_{ab} 
\end{pmatrix} \,.
\end{equation}
that assembles the occupied-occupied, occupied-unoccupied, and unoccupied-unoccupied blocks.

Combining Eqs.~\ref{eq:dgammabydx} and \ref{eq:generalized_formula}, we obtain,
\begin{equation}
 \mathbf{F}^{\text{rSE}}_{A}= - \sum_{p,q} \Lambda_{pq} \sum_{\zeta= \{c, \epsilon , C, V , V^\text{xc}\}} \text{Tr}\left[\dfrac{df_{pq}}{d\zeta} \dfrac{d\zeta}{d\mathbf{R}_A} \right]\, .
 \label{eq:rSE_total_force}
\end{equation}
Thus, the rest of the mission is to examine each of the five individual terms of $\dfrac{f_{pq}}{d\zeta}$ and $\dfrac{d\zeta}{d\mathbf{R}_A}$ for $\zeta =c, \epsilon, C, V,V^\text{xc}$.
\begin{enumerate}
\item $\zeta = c$ (where $c$ is the KS eigenvector).
\begin{equation}
   \text{Tr}\left[\dfrac{f_{pq}}{d\zeta} \dfrac{d\zeta}{d\mathbf{R}_A} \right]_{\zeta=c} 
   =\sum_{i,s} \dfrac{df_{pq}}{dc_{i,s}} \dfrac{dc_{i,s}}{d\mathbf{R}_A} \, .
\end{equation}
Starting from Eqs.~\ref{eq:O_integrals} and \ref{eq:f_matr_in_KS}, after some derivations (see Appendix \ref{sec:df-dc} for details), we can obtain
\begin{equation}
\begin{split}
    \dfrac{df_{pq}}{dc_{i,s}} =&-\sum_{\mu}\sum_{m}^{occ}\sum_{j}  \left( \delta_{ps} c_{j,m} Q_{ij}^{\mu} + c_{j,p} \delta_{ms}  Q_{ji}^{\mu}\right) O_{mq}^{\mu}  \\& 
- \sum_{\mu}\sum_{m}^{occ} \sum_{j} \left( \delta_{ms} c_{j,q} Q_{ij}^{\mu} + c_{j,m} \delta_{qs} Q_{ji}^{\mu} \right)  O_{pm}^{\mu} \\& 
- \sum_{j}\left(\delta_{ps}c_{j,q}V^\text{xc}_{ij}+c_{j,p}\delta_{qs} V^\text{xc}_{ji} \right) \, .
\end{split}
\label{eq:dfbydc}
\end{equation}

In addition, we need to evaluate $\dfrac{dc_{i,s}}{d\mathbf{R}_A}$, the derivative of the KS eigenvectors with respect to the atomic displacement. Here, similar to the approach adopted in Refs.~\cite{Muhammad_2022,Muhammad_2024}, we use the DFPT \cite{RevModPhys.73.515}, as previously implemented in FHI-aims \cite{Shang/etal:2017},  to compute this quantity. In this approach, the first-order change of the KS eigenvector is expanded in terms of
their zeroth-order counterparts, i.e.,
\begin{equation}
   \dfrac{dc_{i,s}}{d\mathbf{R}_A} = c^{(1)}_{i,s} = \sum_r c_{i,r} U_{rs}^{(1)}\, ,
   \label{eq:dcdR_DFPT}
\end{equation}
where $U_{rs}^{(1)}$ are the expansion coefficients that need to be determined. Following the standard procedure, it is straightforward to show that
   \begin{equation} \label{eq:DFPT_U_matrix}
    U_{rs}^{(1)}=
    \begin{cases} \displaystyle
     \sum_{ij} c_{i,r}\dfrac{\epsilon_{s} S^{(1)}_{ij}-H^{(1)}_{ij} }{\epsilon_{r}-\epsilon_{s}} c_{j,s}~, ~~ r\ne s \\
     -\dfrac{1}{2}\sum_{ij}c_{i,r}S^{(1)}_{ij}c_{j,r}~, ~~~~~~~~~~~ r=s
    \end{cases}
    \end{equation}
where $S^{(1)}_{ij} = \dfrac{dS_{ij}}{d\mathbf{R}_A} $ and $H^{(1)}_{ij} = \dfrac{dH_{ij}}{d\mathbf{R}_A} $
are the first-order derivatives (with respect to the atomic displacement) of the overlap and Hamiltonian matrix in the AO representation. These objects are already available in the standard DFPT implementation \cite{Shang/etal:2017}, and hence $U_{rs}^{(1)}$ can be readily computed using Eq.~\ref{eq:DFPT_U_matrix}.

Combining Eqs.~\ref{eq:rSE_total_force}, \ref{eq:dfbydc}, and \ref{eq:dcdR_DFPT}, one finally attains the first 
part of the rSE force, due to the response of the KS eigenvectors to the atomic displacement,
\begin{equation}
 \begin{split}
 \mathbf{F}^{\text{rSE,(1)}}_{A}  = & - \sum_{p,q} \Lambda_{pq} \sum_{i,s} \dfrac{df_{pq}}{dc_{i,s}} \dfrac{dc_{i,s}}{d\mathbf{R}_A} \, \nonumber \\
   = & \sum_{s,r,q} \Lambda_{sq}\Bigg[ \sum_{\mu}\sum_{m}^{occ} O^{\mu}_{rm}O^{\mu}_{mq} + V^\text{xc}_{rq} \Bigg]  U^{(1)}_{rs}   +\\&
     \sum_{r,s,p}  \Lambda_{ps} \Bigg[\sum_{\mu} \sum_{m}^{occ} O^{\mu}_{pm}O^{\mu}_{mr}  + V^\text{xc}_{pr} \Bigg] U^{(1)}_{rs} +\\& 
      \sum_{r,s,p,q} \sum_{\mu}\sum_{m}^{occ}\delta_{ms} \Lambda_{pq}\Bigg [O^{\mu}_{pr}O^{\mu}_{mq} +O^{\mu}_{pm}O^{\mu}_{rq} \Bigg]  U^{(1)}_{rs} \\ 
     = & \sum_{r,s} {\cal G}_{sr}  U^{(1)}_{rs}   
 \end{split}
 \label{eq:rSE_force_firstpart}
\end{equation}
where
 \begin{equation}
    \begin{split} \label{eq:final_G}
        {\cal G}_{s,r}=& \sum_{q}({\bf {\Lambda }}+{\bf{\Lambda} }^{T})_{sq}  \Bigg(\sum_{m}^{occ}\sum_{\mu}  O^{\mu}_{rm}O^{\mu}_{mq} 
        +V^\text{xc}_{rq}\Bigg) \\& 
        +\sum_{m}^{occ}\sum_{\mu} \sum_{p,q} \delta_{sm} O^{\mu}_{mq}({\bf {\Lambda} }+{\bf {\Lambda} }^{T})_{qp}  O^{\mu}_{pr}  \, .
    \end{split}
\end{equation}

Note that the dependence of the force on the atomic moves is through $U^{(1)}_{rs}$, which explicitly depends on 
$\mathbf{R}_A$.

\item $\zeta = \epsilon$ (where $\epsilon$ is the KS eigenvalue).
\begin{equation}
   \text{Tr}\left[\dfrac{f_{pq}}{d\zeta} \dfrac{d\zeta}{d\mathbf{R}_A} \right]_{\zeta=\epsilon} 
   =\sum_{s} \dfrac{df_{pq}}{d\epsilon_{s}} \dfrac{d\epsilon_{s}}{d\mathbf{R}_A} \, .
\end{equation}
From Eq.~\ref{eq:f_matr_in_KS}, it is easy to see
\begin{equation}
    \dfrac{df_{pq}}{d\epsilon_{s}} = \delta_{pq}\delta_{ps}
\label{eq:dfbyde}
\end{equation}
The derivative of the KS eigenvalues with respect to the atomic displacement can also be obtained from DFPT
\cite{Shang/etal:2017,Muhammad_2022},
  \begin{equation}
  \dfrac{d\epsilon_{s}}{d\mathbf{R}_A} = \epsilon_{s}^{(1)}=\sum_{ij}c_{i,s}\left[H^{(1)}_{ij}-\epsilon_{s}^{(0)} S^{(1)}_{ij}\right]c_{j,s}\, .
   \end{equation}
   Thus, the second part of the rSE force is rather simple,
 \begin{equation}
 \begin{split}
 \mathbf{F}^{\text{rSE,(2)}}_{A}  = & - \sum_{pq} \Lambda_{pq} \sum_{s} \dfrac{df_{pq}}{d\epsilon_{s}} \dfrac{d\epsilon_{s}}{d\mathbf{R}_A} \,  \\ &
 = -\sum_{s} \Lambda_{ss} \epsilon^{(1)}_s \, .
 \end{split}
 \end{equation}
 Again, the dependence of this part of the force on the atomic position is via $\epsilon^{(1)}_s$.

\item $\zeta = C$ (where $C$ is the RI expansion coefficient).
Expressing the trace operation of this term explicitly, we have
\begin{equation}
   \text{Tr}\left[\dfrac{f_{pq}}{d\zeta} \dfrac{d\zeta}{d\mathbf{R}_A} \right]_{\zeta=C} 
   =\sum_{i,j,\mu} \dfrac{df_{pq}}{dC_{ij}^{\mu}} \dfrac{dC_{ij}^{\mu}}{d\mathbf{R}_A} \, .
\end{equation}

Starting from Eqs.~\ref{eq:f_matr_in_KS}, \ref{eq:O_integrals}, and \ref{eq:Q_integrals}, we can obtain
 \begin{equation}
    \begin{split}
        \dfrac{df_{pq}}{dC^{\mu}_{ij}} =& -\sum_{\nu}\sum_{m}^{occ} \dfrac{d}{dC^{\mu}_{ij}} 
         \Bigg ( O^{\nu}_{pm} O^{\nu}_{mq} \Bigg )  \\
        = &-\sum_{\nu\nu'}\sum_{m}^{occ}\sum_{kl} V^{1/2}_{\nu\nu'}
        \Bigg (  \delta_{ik} \delta_{jl} \delta_{\mu \nu'} c_{kp}c_{lm} O^{\nu}_{mq}  +  \delta_{ik} \delta_{jl} \delta_{\mu \nu'} c_{km}c_{lq} O^{\nu}_{pm}  \Bigg)\\
        = & -\sum_{\nu}\sum_{m}^{occ} V_{\nu \mu}^{1/2} \Bigg ( c_{ip}c_{jm} O^{\nu}_{mq}+c_{im} c_{jq} O^{\nu}_{pm}  \Bigg )\, .
    \end{split} 
\end{equation}
 Now, we are still left with the computation of $\dfrac{dC_{ij}^{\mu}}{d\mathbf{R}_A}$. Within the local RI approximation, the RI coefficients $C_{ij}^{\mu}$ are reduced to a two-center quantity \cite{Ihrig/etal:2015,Lin/Ren/He:2020}. Namely, suppose the AOs $i$ and $j$ belong to two atoms $I$ and $J$, $C_{ij}^{\mu}$ are  non-zero only if the ABF $\mu$ is centering on one of them. Obviously, $\dfrac{dC_{ij}^{\mu}}{d\mathbf{R}_A}$ is only non-zero if the atom $A$ is either $I$ or $J$. In this case, as discussed in Ref.~\cite{Muhammad_2022}, we have
 \begin{align} \label{eq:C_derivative} \displaystyle
 \dfrac{dC^{\mu}_{ij}}{d\mathbf{R}_A}=\begin{cases}
 \dfrac{dC^{\mu}_{ij}}{d\mathbf{R}_{IJ}}\, ,  \quad\quad\quad  A=I\\ \\ 
 -\dfrac{dC^{\mu}_{ij}}{d\mathbf{R}_{IJ}}\, ,  \quad\quad A=J\\  \\
 0\, , \quad\quad\quad\quad\quad\quad \text{else}
 \end{cases}
\end{align}
where $\mathbf{R}_{IJ}$ is the spatial vector directing from $J$ to $I$. $\dfrac{dC^{\mu}_{ij}}{d\mathbf{R}_{IJ}}$ are readily available in the local RI infrastructure as implemented in FHI-aims \cite{Ihrig/etal:2015}. Putting these ingredients together, the third part of the rSE force is given by

 \begin{equation} \label{eq:final_force_333}
 \begin{split}
  \mathbf{F}^{\text{rSE,(3)}}_{A}  =&  - \sum_{pq} \Lambda_{pq} \sum_{i,j,\mu} \dfrac{df_{pq}}{dC_{ij}^\mu} \dfrac{dC_{ij}^\mu}{d\mathbf{R}_A} \,  \\ 
 =& \sum_{pq} \Lambda_{pq} \sum_{\nu}\sum_{m}^{occ} 
   \left[ \sum_{i \in A,j,\mu}  V_{\nu \mu}^{1/2} \Bigg ( c_{ip}c_{jm} O^{\nu}_{mq}+c_{im} c_{jq} O^{\nu}_{pm}  \Bigg ) \right. \dfrac{dC^{\mu}_{ij}}{d\mathbf{R}_{IJ}} \\ &
   ~~~~- \left. \sum_{j \in A,i,\mu}  V_{\nu \mu}^{1/2} \Bigg ( c_{ip}c_{jm} O^{\nu}_{mq}+c_{im} c_{jq} O^{\nu}_{pm}  \Bigg ) \dfrac{dC^{\mu}_{ij}}{d\mathbf{R}_{IJ}} \right]\, ,
 \end{split}
 \end{equation}
 where $i \in A$ means the summation only goes over the AOs that are located on the atom $A$. Here, one should keep in mind that, in the
 above two lines, $A=I$ and $A=J$, respectively.
 
\item $\zeta = V$ (where $V$ is the Coulomb matrix).
The fourth term of the rSE force is given by
\begin{equation}
   \text{Tr}\left[\dfrac{f_{pq}}{d\zeta} \dfrac{d\zeta}{d\mathbf{R}_A} \right]_{\zeta=V} 
   =\sum_{\mu,\nu} \dfrac{df_{pq}}{dV_{\mu\nu}} \dfrac{dV_{\mu\nu}}{d\mathbf{R}_A} \, .
\end{equation}
Again, starting from Eqs.~\ref{eq:f_matr_in_KS}, \ref{eq:O_integrals}, and \ref{eq:Q_integrals}, after a few lines of
derivations (see Appendix \ref{sec:df-dv} for further details), it can be shown that
\begin{equation}
    \begin{split}
\dfrac{df_{pq}}{dV_{\mu \nu}}&= - \sum_{m}^{occ} \sum_{\mu' \nu'}V^{-\frac{1}{2}}_{\mu \mu'} O^{\mu'}_{pm}  O^{\nu'}_{mq}V^{-\frac{1}{2}}_{\nu' \nu}\, ,
    \end{split}
    \label{eq:dfbydV}
\end{equation}
As for the variation of the Coulomb matrix with respect to the atomic displacement, similar to the RI expansion coefficients, one has
\begin{align} \label{eq:V_UV_derivative} \displaystyle
 \dfrac{dV_{\mu\nu}}{d\mathbf{R}_A}=\begin{cases}
 \dfrac{dV_{\mu\nu}}{d\mathbf{R}_{\cal UV}}\, ,  \quad\quad\quad  A={\cal U} \, .\\ \\ 
 -\dfrac{dV_{\mu\nu}}{d\mathbf{R}_{\cal UV}}\, ,  \quad\quad A={\cal V} \, .\\  \\
 0\, , \quad\quad\quad\quad\quad\quad \text{else}\, .
 \end{cases}
\end{align}
where ${\cal U}$ and ${\cal V}$ denote the atoms on which the ABFs $\mu,\nu$ are centered, and $\mathbf{R}_{\cal UV}$
is the spatial vector directing from the atom ${\cal V}$ to ${\cal U}$. Combining Eqs.~\ref{eq:rSE_total_force},
\ref{eq:dfbydV} and \ref{eq:V_UV_derivative}, we obtain the explicit expression of the fourth part of the rSE force,
\begin{equation}
     \begin{split}
 \mathbf{F}^{\text{rSE,(4)}}_{A} &=    - \sum_{pq} \Lambda_{pq} \sum_{\mu,\nu} \dfrac{df_{pq}} {dV_{\mu\nu}} \dfrac{dV_{\mu\nu}^\mu}{d\mathbf{R}_A}    \\&
 = \sum_{pq} \Lambda_{pq} \sum_{m}^{occ} \left[\sum_{\mu \in A, \nu} \sum_{\mu' \nu'}V^{-\frac{1}{2}}_{\mu \mu'} O^{\mu'}_{pm}  O^{\nu'}_{mq}V^{-\frac{1}{2}}_{\nu' \nu} \dfrac{dV_{\mu\nu}}{d\mathbf{R}_{\cal UV}}
  -  \sum_{\mu, \nu \in A} \sum_{\mu' \nu'}V^{-\frac{1}{2}}_{\mu \mu'} O^{\mu'}_{pm}  O^{\nu'}_{mq}V^{-\frac{1}{2}}_{\nu' \nu} \dfrac{dV_{\mu\nu}}{d\mathbf{R}_{\cal UV}} \right]\, .
 \end{split}
\end{equation}
For the two terms in the bracket, $A={\cal U}$ and ${\cal V}$, respectively.

\item $\zeta = V^\text{xc}$ (where $V^{xc}$ is the XC potential matrix).
This term is given explicitly as
\begin{equation}
   \text{Tr}\left[\dfrac{f_{pq}}{d\zeta} \dfrac{d\zeta}{d\mathbf{R}_A} \right]_{\zeta=V^\text{xc}} 
   =\sum_{ij} \dfrac{df_{pq}}{dV^\text{xc}_{ij}} \dfrac{dV^\text{xc}_{ij}}{d\mathbf{R}_A} \, .
\end{equation}
Based on Eqs.~\ref{eq:f_matr_in_KS} and \ref{eq:O_integrals}, one can immediately see that
\begin{equation}
  \dfrac{df_{pq}}{dV_{ij}^\text{xc}}= - c_{i,p}^\ast c_{j,q} \, .
 \label{eq:dfbydVxc}
\end{equation}
Thus, we arrive at the following expression for the fifth part of the rSE force
\begin{equation}
    \mathbf{F}^{\text{rSE,(5)}}_{A} = \sum_{pq} \Lambda_{pq} \sum_{ij}c_{i,p}^\ast c_{j,q} \dfrac{dV^\text{xc}_{ij}}{d\mathbf{R}_A} \, .
\end{equation}
Here, $\dfrac{dV^\text{xc}_{ij}}{d\mathbf{R}_A}$ can be computed in a similar way as the first order Hamiltonian $H^{(1)}_{ij}$.
Formally, 
\begin{equation}
\begin{split}
    \dfrac{dV^{xc}_{ij}}{d\textbf{R}_{A}} =&  \int \dfrac{d\phi_{i}(\textbf{r})}{d\textbf{R}_{A}} v_{xc}(\textbf{r}) \phi_{j}(\textbf{r}) d\textbf{r} +  \int \phi_{i}(\textbf{r}) \dfrac{dv_{xc}(\textbf{r})}{d\textbf{R}_{A}}  \phi_{j}(\textbf{r}) d\textbf{r}+ 
    \int \phi_{i}(\textbf{r})  v_{xc}(\textbf{r})  \dfrac{d\phi_{j}(\textbf{r})}{d\textbf{R}_{A}}
    \end{split}
\end{equation}
and the integration can be done via quadrature on a 3D real-space grid.

\end{enumerate}

So far, we have discussed thoroughly how the five terms of the rSE force are computed. In our practical implementation, the rSE force can be calculated together with the RPA force, in the following sense,
\begin{equation}
 \mathbf{F}^{\text{RPA+rSE}}_{A} = - \sum_{\zeta= \{c, \epsilon , C, V , V^\text{xc}\}} \text{Tr}\left[\left(\dfrac{E_c^\text{RPA}}{d\zeta}+\dfrac{E_c^\text{rSE}}{d\zeta}\right) \dfrac{d\zeta}{d\mathbf{R}_A} \right]\, .
 \label{eq:RPA+rSE_total_force}
\end{equation}
According to Eq.~\ref{eq:RPA+rSE_total_force}, we just need to evaluate $\dfrac{E_c^\text{rSE}}{d\zeta}$
for each intermediate variable $\zeta= \{c, \epsilon , C, V , V^\text{xc}\}$, and add them to the corresponding
terms of the RPA correlation energy. The final step, i.e., the contraction with $\dfrac{d\zeta}{d\mathbf{R}_A}$,
can be done together for RPA and rSE parts. The only caveat is that when $\zeta= V^\text{xc}$, the RPA part of the force is zero, and we are left with only the rSE part.

\section{Implementation and Computational Details}
\label{sec:imple}
The rSE analytical gradient formalism described in the previous section has been implemented in the FHI-aims code package \cite{Blum/etal:2009,Havu/etal:2009,Ren/etal:2012}. The parallel efficiency and computation time of the RPA+rSE gradients are comparable to the standard RPA case, since the time consumption of the rSE force evaluation  only amounts to a small fraction of the total cost. In Algorithm~\ref{alg:rSE_force_algorithm}, 
we show the major execution steps in the computation of the first and third parts 
of the rSE force, where the formal scaling of each step is given in parentheses on the right side. One can see that the highest scaling among all the steps is formally $\mathcal{O}(N^4)$. However, the prefactor of the $\mathcal{O}(N^4)$ steps is rather small, in particular because, unlike the RPA case, there is no frequency dependence involved in the rSE calculations. In addition to the first and third parts, the fourth part of the rSE force also has a formal $\mathcal{O}(N^4)$ computational scaling. The corresponding workflow of the computational algorithm for this part can be designed similarly, in analogy to what is presented in Algorithm~\ref{alg:rSE_force_algorithm}. To save space, this will not be discussed in further detail here. 

\begin{figure}[H]
	\centering
		\includegraphics[width=1.0\textwidth]{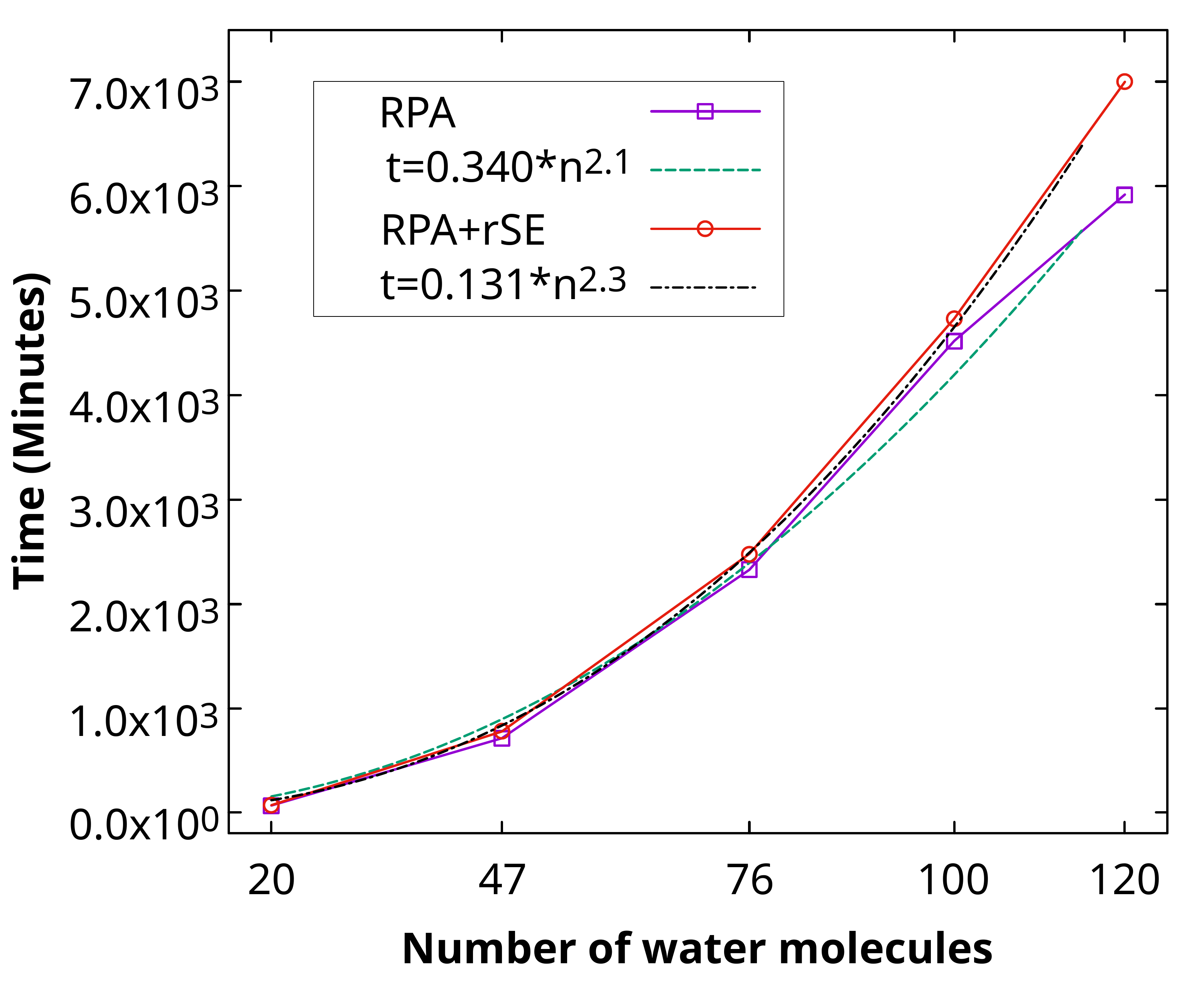} 
	\caption{ \label{fig:scaling_plot}Wall-clock timings (Minutes) of one RPA and RPA+rSE relaxation step as a function of the water cluster size $n$. We have used cc-pVTZ basis sets together with the frozen-core approximation. For these calculations the 320 CPU cores are used. The polynomial fit of the scaling behavior of the computational times is also added to the graph.}
\end{figure}

To check the computational efficiency of our rSE force implementations, we performed RPA and RPA+rSE force calculations for water clusters (H$_2$O)$_n$ up to $n=120$. The computation times are plotted in Fig.~\ref{fig:scaling_plot} for cluster sizes $n=20$, 47, 76, 100, and 120. 
In recent works, in addition to the rSE force implementation, we also refined our RPA force implementation, resulting in noticeable performance improvement over previous versions \cite{Muhammad_2022, Muhammad_2024}. Figure~\ref{fig:scaling_plot} shows that the rSE correction to the force calculation only adds marginal extra cost compared to the standard RPA.
A polynomial regression of the timing data yields scaling behaviors well described by the relation 
$t(n)=bn^{\alpha}$, where the scaling exponents are $\alpha=2.1$ for RPA and 
$\alpha=2.3$ for RPA+rSE. 
These results demonstrate that, despite the formal  
$\mathcal{O}(N^{4})$ scaling, the actual computational cost of both RPA and RPA+rSE force calculations grows nearly quadratically with system size up to 100 water molecules. This practical performance significantly improves the feasibility of applying RPA and RPA+rSE approaches to systems of high physical interest.

\begin{figure*} 
	\begin{minipage}{1.0\linewidth}
		\begin{algorithm}[H]
 \begin{center}
     \caption{Flowchart for evaluating the first and third parts of the rSE force (Eq.~\ref{eq:final_force_333}). Here $N_{at}$, $N_{b}$, $N_{occ}$, $N_{unocc}$, $N_{aux}$, $N_{pairs}$ are the numbers of atoms, the AO basis functions, the occupied states, the unoccupied states, the ABFs, and the numbers of atom pairs, respectively. In practical calculations, $N_{aux}>N_{b}=N_{occ}+N_{unocc}$. The formal scaling of the major steps are indicated on the right. For the lines which involve more than one execution steps, only the scaling of the most expensive step is given.} \label{alg:rSE_force_algorithm}
      \end{center}
			\begin{algorithmic}[1]
				\vspace{-16.pt}
				\Statex \hspace{8.5cm}	\textbf{Scaling Behavior}
				\vspace{4pt}
                \State Construct $O^{\nu}_{mn}$ ~~ (cf. Eq.~\ref{eq:O_integrals}) ~~~~~~~~~~~~~~~~~~~~~~~~~~$\mathcal{O}(N_{aux}N_{b}^2(N_{occ}+N_{unocc}))$
				\vspace{4pt}
             \State Calculate $f_{pq}$ ~~~~~
             (cf. Eq.~\ref{eq:f_matr_in_KS}) ~~~~~~~~~~~~~~~~~~~~~~~~~~~~ $\mathcal{O}(N_{aux}N_{occ}N_{b}^{2})$
             \vspace{4pt} 
             \State Calculate $F_{pq}$ ~~~~~ (cf. Eq.~\ref{eq:Fou}) ~~~~~~~~~~~~~~~~~~~~~~~~~~~~~$\mathcal{O}(N^{3}_{b})$
                \vspace{4pt}
				\State Construct $\Lambda_{pq}$  ~~~~~(cf. Eq.~\ref{eq:occ_occ_block}, \ref{eq:unocc_unocc_block}, \ref{eq:occ_unocc_block}) ~~~~~~~~~~~~~~~~~~~$\mathcal{O}(N_{b}^3)$
				\vspace{8pt}
				\State ${\cal G}_{sr}=0\,; \quad \textbf{F}^{\text{rSE},(1)}=0$
				\vspace{6pt}
                \State $D_{rq}\gets \sum_{m}^{occ}\sum_{\mu} O^{\mu}_{rm}O^{\mu}_{mq}$ ~~~~~(cf. Eq.~\ref{eq:final_G})   ~~~~~~~~~~~~~~$\mathcal{O}(N_{aux}N_{b}^{2}N_{occ})$  
                \vspace{6pt}
                \State ${\cal G}_{sr} \gets \sum_{q} (\Lambda +\Lambda^{T})_{sq} (D+V^{xc} )_{rq}$ ~~~(cf. Eq.~\ref{eq:final_G})  ~~~~~~$\mathcal{O}(N_{b}^{3})$ 
                \vspace{6pt} 
                \State ${\cal Y}_{qr}^{\mu} \gets \sum_{p}(\Lambda +\Lambda^{T})_{qp} O^{\mu}_{pr}$ ~~~~~cf. Eq.~\ref{eq:final_G})   ~~~~~~~~~~~~~~~~$\mathcal{O}(N_{aux}N_{b}^{3})$ 
                \vspace{6pt}  
                \State ${\cal G}_{mr} \gets {\cal G}_{mr} +\sum_{\mu}\sum_{p} O^{\mu}_{mq} {\cal Y}^{\mu}_{qr}$ ~~~~~cf. Eq.~\ref{eq:final_G}) ~~~~~~~~~$\mathcal{O}(N_{aux}N_{b}^{2}N_{occ} )$
                \vspace{6pt}
                \State $\textbf{F}^{\text{rSE},(1)} \gets \sum_{sr}{\cal G}_{sr} U^{(1)}_{rs}$ ~~~~~~~~~~~~~~~~~~~~~~~~~~~~~~~~~~~~~~~$\mathcal{O}(N_{b}^{3})$ 
                \vspace{4pt}
                \State $\textbf{F}^{\text{rSE, }(3)}=0$ 
                \vspace{4pt}
                \State ${\cal C}^{\nu}_{mq} \gets \sum_{\mu} V^{\frac{1}{2}}_{\nu \mu} O^{\mu}_{mq} $
                 \vspace{8pt}  ~~~~(cf. Eq.~\ref{eq:final_force_333}) ~~~~~~~~~~~~~~~~~~~~~~~$\mathcal{O}(N_{aux}^{2}N_{b}N_{occ})$ 
                 \vspace{4pt}
                 \State ${\cal C}^{\nu}_{pm} \gets \sum_{q} {\cal C}^{\nu}_{mq} (\Lambda +\Lambda^{T})_{pq}$  ~~~~(cf. Eq.~\ref{eq:final_force_333}) ~~~~~~~~~~~~~~~$\mathcal{O}(N_{aux}N_{b}^{2}N_{occ})$ 
                 \vspace{8pt} 
                 \State ${\cal C}^{\nu}_{im} \gets \sum_{p}c_{ip} {\cal C}^{\nu}_{pm}$ ~~~~(cf. Eq.~\ref{eq:final_force_333}) ~~~~~~~~~~~~~~~~~~~~~~~~~~~$\mathcal{O}(N_{aux}N_{b}^{2}N_{occ})$ 
                 \vspace{4pt}
                 
              \For {${\cal K} \gets 1$ to $N_{-}pairs$}
				\vspace{4pt}
                \State  ${\cal I}$: first atom in pair, $\cal J:$ second atom in pair, ${\cal I}>{\cal J}\,, \quad \forall (i\in {\cal I}, j\in {\cal J}) $
                \vspace{4pt}
                \State Compute $\dfrac{dC^{\nu}_{i, j}}{dR_{\cal I \cal J}}$ \quad  $\forall ~(\nu \in {\cal I})$ 
                \vspace{4pt} 
                \State ${\cal C}^{\nu}_{ij} \gets \sum_{m}^{occ}c_{i, m} {\cal C}^{\nu}_{j,m} +\sum_{m}^{occ} {\cal C}^{\nu}_{i,m} c_{j, m}$  
				\vspace{4pt}
                \State $\textbf{F}^{\text{rSE, }(3)}_{\cal I} \gets \textbf{F}^{\text{rSE, }(3)}_{\cal I} -2\sum_{ij}\sum_{\nu} C^{\nu}_{ij}\dfrac{dC^{\nu}_{i, j}}{dR_{\cal I \cal J}}$ ~~~(cf. Eq.~\ref{eq:final_force_333}) ~~~~$\mathcal{O}(3N_{at}N_{aux}N_{b}^{2})$ 
                \vspace{4pt}
                \State $\textbf{F}^{\text{rSE, }(3)}_{\cal J} \gets \textbf{F}^{\text{rSE, }(3)}_{\cal J} +2\sum_{ij}\sum_{\nu} C^{\nu}_{ij}\dfrac{dC^{\nu}_{i, j}}{dR_{\cal I \cal J}}$ ~~~(cf. Eq.~\ref{eq:final_force_333}) ~~~~$\mathcal{O}(3N_{at}N_{aux}N_{b}^{2})$
                \vspace{4pt}
                \State Compute $\dfrac{dC^{\nu}_{i, j}}{dR_{\cal I \cal J}}$ \quad  $\forall ~(\nu \in {\cal J})$ 
                \vspace{4pt}
                 \State ${\cal C}^{\nu}_{ij} \gets \sum_{m}^{occ}c_{i, m} {\cal C}^{\nu}_{j,m} +\sum_{m}^{occ} {\cal C}^{\nu}_{i,m} c_{j, m}$   
				\vspace{4pt}
                \State $\textbf{F}^{\text{rSE, }(3)}_{\cal I} \gets \textbf{F}^{\text{rSE, }(3)}_{\cal I} -2\sum_{ij}\sum_{\nu} C^{\nu}_{ij}\dfrac{dC^{\nu}_{i, j}}{dR_{\cal I \cal J}}$ ~~~(cf. Eq.~\ref{eq:final_force_333}) ~~~~$\mathcal{O}(3N_{at}N_{aux}N_{b}^{2})$
                \vspace{4pt}
                \State $\textbf{F}^{\text{rSE, }(3)}_{\cal J} \gets \textbf{F}^{\text{rSE, }(3)}_{\cal J} +2\sum_{ij}\sum_{\nu} C^{\nu}_{ij}\dfrac{dC^{\nu}_{i, j}}{dR_{\cal I \cal J}}$~~~(cf. Eq.~\ref{eq:final_force_333}) ~~~~~~$\mathcal{O}(3N_{at}N_{aux}N_{b}^{2})$
			\EndFor
			\end{algorithmic} 
		\end{algorithm}
	\end{minipage}
\end{figure*}

In FHI-aims, both numerical atomic orbitals (NAOs) and Gaussian-type orbitals (GTOs) are supported as basis functions for RPA+rSE gradient computations. For any set of AO basis functions, the ABFs are generated ``on the fly'' following the automatic procedure described in Refs.~\citenum{Ren/etal:2012,Ihrig/etal:2015}, with very tight thresholds. The actual AO basis sets used in the calculations will be specified later in the results part. A modified Gauss-Legendre frequency grid consisting of 24 points is used for frequency integration in Eq.~\ref{eq:EcRPA_RI}. 
The Perdew-Burke-Ernzerhof (PBE) \cite{Perdew/Burke/Ernzerhof:1996} generalized gradient approximation (GGA) is used to generate the orbitals and orbital energies used in the RPA and rSE calculations.

\section{Results and discussion}
\label{sec:results}
In this section, we will present the major results obtained in this work. We start by checking the correctness of our implementation by comparing the analytical RPA+rSE forces and their counterparts obtained by the finite difference (FD) method. We then present the bond lengths and bond angles of a test set of small molecules to demonstrate the performance of the RPA+rSE method in determining geometries. 
We then proceed to relax the geometries of water clusters by RPA+rSE, and examine the influence of the geometries on the energetics.

\subsection{Validation check}
\label{validity}
To verify the correctness of our implementation, we compare the RPA+rSE forces obtained using the analytical gradient and  FD approaches for a set of diatomic molecules at fixed bond lengths. The NAO basis set is used in these calculations, with the standard ``tight'' settings of FHI-aims for the basis size, cutoff radius, and numerical integration grid. The FD force is calculated as follows:
\begin{equation}
    F_{\text{FD}}(x)= \dfrac{E^{\text{RPA+rSE}}(x+h)-E^{\text{RPA+rSE}}(x-h)}{2h}
\end{equation}
where $x$ denotes the bond length of the molecule at which the force is to be determined, and the displacement $h$ is chosen to be 0.001 {\AA}. In Table~\ref{tab:finite_difference_tight_default}, we present the analytical and FD RPA+rSE forces for a set of diatomic molecules. To highlight the contribution from rSE, the rSE component of the force is also presented separately on the right side of Table~\ref{tab:finite_difference_tight_default}.
Table~\ref{tab:finite_difference_tight_default} clearly demonstrates excellent agreement between analytical and
FD results for both the total RPA+rSE force and the rSE component.  For all molecules, the differences are approximately 1 meV/\AA~ for the RPA+rSE forces, and 0.1 meV/\AA~ or even smaller for the rSE part. A relative deviation as low as $0.41\%$ is observed for RPA+rSE and $0.20\%$ for rSE, which is completely within acceptable error tolerance. In particular, the bond lengths of the molecules listed in Table~\ref{tab:finite_difference_tight_default} were chosen somewhat arbitrarily. As a result, the magnitude of the forces varies widely. However, the analytical RPA+rSE forces and the FD forces always agree to a high precision.  Note that we don’t expect perfect agreement anyway since the FD results also include higher-order effects. In brief, this benchmark clearly demonstrates the validity of our RPA+rSE analytical gradient implementation.

\begin{table*} [ht]
	\caption{\label{tab:finite_difference_tight_default}The full (RPA+rSE)@PBE forces and its rSE contributions for a set of diatomic molecules 
		calculated using the analytical gradient (labeled as $F$, in eV/\AA ) and finite difference (FD) methods (labeled as $F_\text{FD}$). The ``\textit{tight}" NAO basis set is used in these calculations. 
		The ``Dist" column lists the bond lengths of the molecules at which the forces are calculated.  For each molecule, the difference $\Delta F=F-F_\text{FD}$ and absolute percentage 
		deviation (APD), given by $|\Delta F|/F_\text{FD}*100\%$, are also shown. The mean absolute percentage deviation (MAPD) is given by the average of APD over all molecules.}
	\centering
	\scalebox{0.75}{
		\begin{tabular}{lcrrrrrrrrr} 
			\hline\hline 
			\multicolumn{1}{c}{\multirow{2}{*}{System}} & \multicolumn{1}{c}{\multirow{2}{*}{Dist.(\AA)}} & \multicolumn{4}{c}{(RPA+rSE)@PBE} & & \multicolumn{4}{c}{rSE@PBE}  \\
			\cline{3-6} \cline{8-11}
			&  &\multicolumn{1}{c}{\multirow{1}{*}{$F$ (eV/\AA)}} & \multicolumn{1}{c}{\multirow{1}{*}{$F_\text{FD}$ (eV/\AA)}} & \multicolumn{1}{c}{\multirow{1}{*}{$\bigtriangleup F$ (eV/\AA)}}& \multicolumn{1}{c}{\multirow{1}{*}{ APD }} & &  \multicolumn{1}{c}{\multirow{1}{*}{$F$ (eV/\AA)}} & \multicolumn{1}{c}{\multirow{1}{*}{$F_\text{FD}$ (eV/\AA)}} & \multicolumn{1}{c}{\multirow{1}{*}{$\bigtriangleup F$ (eV/\AA)}}& \multicolumn{1}{c}{\multirow{1}{*}{ APD }} \\ 
			\hline 
			H$_{2}$  & 0.7496  & 0.144725 &  0.143980 &  0.000745 &   0.52\% & & 0.035087 & 0.035095  & $-$0.000008 & 0.02\% \\
			Li$_{2}$  & 2.6257 & 0.116125 &  0.118490  & $-$0.002365 &   1.99\% & & 0.014054 & 0.014070 & $-$0.000016 & 0.11\% \\ 
			Be$_{2}$  & 2.5000 & 0.078119 & 0.077460 &  0.000659 &   0.85\% & & 0.007541 &0.007555 & $-$0.000014  & 0.19\% \\
			N$_{2}$  & 1.1220  & 1.286060 & 1.287100 & $-$0.001040 &  0.08\%  & & 0.611784 & 0.611880 & $-$0.000096 & 0.02\% \\
			F$_{2}$  & 1.0000  & 58.218000 & 58.220600 & $-$0.002600 &  0.01\% & & 0.083826 &  0.083650 & 0.000176 & 0.21\%\\
			Na$_{2}$   &3.0790 & 0.097150 & 0.098510 & $-$0.001360 &  1.38\% & & 0.015057 & 0.014895 & 0.000162 & 1.09\% \\
			HF  & 1.0000   & 3.458120 & 3.456940 & 0.001180 &   0.03\% & & 0.175665 &  0.175695 & $-$0.000030 & 0.02\% \\
			BF  & 1.2570   & 1.017660  & 1.020270  & $-$0.002610 &  0.26\% & & 0.454789 & 0.454760  & 0.000029 & 0.01\% \\
			CO  & 1.0000   & 26.445400 &  26.445000 & 0.000400 &  0.00\% & & 0.563524  & 0.563620  & $-$0.000096 & 0.02\%\\
			$\text{NO}^{+}$  & 1.0656  & 2.066140 & 2.068410 & $-$0.002270 &  0.11\% &  & 0.927389 &  0.927465  & $-$0.000076  & 0.01\% \\
			HCl   & 2.0000  & 3.589930  &  3.589950 & $-$0.000020&    0.00\% & & 0.140370 & 0.140350 & 0.000020 & 0.01\%\\
			Cl$_{2}$  & 1.9878  & 0.697251  & 0.696935 & 0.000316 &   0.05\% & & 0.046277 & 0.045965 & 0.000312 & 0.68\%\\ 
			MAE & & & & 0.001297   &  & & & & 0.000086 \\ 
			MAPD  &  & & & &  0.44\% & & & & & 0.20\% \\ 
			\hline\hline
	\end{tabular}}
\end{table*}

\subsection{Optimized RPA+rSE Geometries for Small Molecules}
\label{sec:small_molecules}
With the successful implementation of RPA+rSE analytical gradients in FHI-aims, we can optimize the geometrical configurations of molecular systems with the RPA+rSE method and assess its performance in determining molecular structures. In Table~\ref{tab:comparison_of_results}, we present the optimized equilibrium (RPA+rSE)@PBE bond lengths
for a set of 26 small molecules (and bond angles for the polyatomic molecules). They are compared with the reference geometries and those obtained by PBE, PBE0, RPA@PBE, and MP2 methods for the same set of molecules.  The RPA@PBE results for this set of molecules have been previously reported in Ref.~\citenum{Muhammad_2022}.

\begingroup
\def\arraystretch{0.70}
\setlength\tabcolsep{.1pt}
\setlength\LTcapwidth{.96\linewidth}
    	\begin{longtable}{lcrrrrrrr}
    		\caption{\label{tab:comparison_of_results} (RPA+rSE)@PBE  bond lengths $R_e$ (in pm)  and bond angles $\varTheta_{e}$ (deg) with those
	of PBE, PBE0 and MP2, obtained using cc-pVQZ basis set. The (RPA+rSE)@PBE and RPA@PBE geometries obtained using cc-pV5Z basis sets are also presented.  For all these method the difference ($\Delta D$) from the reference values are presented.
	 For rare gas dimers, additional diffuse functions (i.e., aug-cc-pVQZ and aug-cc-pV5Z) are used in (RPA+rSE)@PBE and RPA@PBE calculations. The reference values are taken from ~\onlinecite{Filip/Poul/Jeppe/stanton/2002,COOK/1975,Tang/Toennies:2003, CCCBDB}, 
     which are indicated with superscript $a$,~ $b$,~ $c$~  and $d$ respectively. The mean error (ME) and mean absolute error (MAE) are given for each method. The mean absolute percentage error (MAPE) given by $|\Delta D|/\text{Reference} \times 100\% $ is also presented. } \\
    \hline\hline 
		\multicolumn{1}{c}{\multirow{2}{*}{System}} &
		\multicolumn{1}{c}{\multirow{2}{*}{~$R_e/ \varTheta_{e}$}} &
		\multicolumn{1}{c}{\multirow{2}{*}{~Ref. \footnotemark[1] }} &
		\multicolumn{1}{c}{\multirow{2}{*}{(RPA+rSE)/QZ}}&
		\multicolumn{1}{c}{\multirow{2}{*}{(RPA+rSE)/5Z}}& 
		\multicolumn{1}{c}{\multirow{2}{*}{RPA/5Z}} &
		 \multicolumn{1}{c}{\multirow{2}{*}{~~PBE}} &  
		 \multicolumn{1}{c}{\multirow{2}{*}{~PBE0}} &
		 \multicolumn{1}{c}{\multirow{2}{*}{MP2\cite{CCCBDB, nwchem}}} \\
		& & & \\ 
\hline
\endfirsthead
\multicolumn{9}{c}%
{\tablename\ \thetable\ -- \textit{Continued from previous page}} \\
\hline\hline 
		\multicolumn{1}{c}{\multirow{2}{*}{System}} &
		\multicolumn{1}{c}{\multirow{2}{*}{~$R_e/ \varTheta_{e}$}} &
		\multicolumn{1}{c}{\multirow{2}{*}{~Ref. \footnotemark[1] }} &
		\multicolumn{1}{c}{\multirow{2}{*}{(RPA+rSE)/QZ}}&
		\multicolumn{1}{c}{\multirow{2}{*}{(RPA+rSE)/5Z}}& 
		\multicolumn{1}{c}{\multirow{2}{*}{RPA/5Z}} &
		 \multicolumn{1}{c}{\multirow{2}{*}{~~PBE}} &  
		 \multicolumn{1}{c}{\multirow{2}{*}{~PBE0}} &
		 \multicolumn{1}{c}{\multirow{2}{*}{MP2\cite{CCCBDB, nwchem}}} \\
		& & & \\ 
\hline
\endhead
\hline \multicolumn{9}{r}{\textit{Continued on next page}} \\
\endfoot
\hline
$^1$\onlinecite{Filip/Poul/Jeppe/stanton/2002} & $^2$\onlinecite{COOK/1975}
		& $^3$\onlinecite{Tang/Toennies:2003} & $^4$\onlinecite{CCCBDB} & & & 
\endlastfoot
\hline  \\ [-0.5ex]
			\multicolumn{7}{c}{Bond lengths} \\ [3ex]
		H$_{2}$ & $R_{e}$ & 74.15~  & 0.24~  &  0.23  & 0.13~  &0.87~ &       0.30  & $-$0.54~   \\

		F$_{2}$ & $R_{e}$& 141.27~  &  3.69~  & 3.52 & 2.18~ &  0.06~ & $-$3.71  & $-$1.56~ \\
		
		N$_{2}$ & $R_{e}$& 109.76~ & 1.23~  &  1.12  & 0.68~ &  0.48~  &  $-$0.86  &  1.28~   \\
		
		HF & $R_{e}$& 91.68~  & 0.78~   & 0.84   & 0.48~ & 1.25~  &  0.05  &  0.04~    \\
		
		CO & $R_{e}$& 112.82~   & 1.67~  & 1.58   & 0.84~  & 0.71~ &	 $-$0.59  & 0.64~   \\
		
		CO$_{2}$ & $R_{e}$& 116.01~   & 1.77~ &  1.68 & 0.83~ & 1.03~  &  $-$0.35   & 0.61~   \\
		
	\multirow{1}{*}{\text{HCN}}   & $R_{e}$ (H-C) &  106.53~  & 0.62~  & 0.57  &  0.32~ & 0.93~   &  0.22   &0.11~  \\ 
		& $R_{e}$ (C-N) & 115.34~ & 1.25~  &  1.14  &  0.65~ & 0.40~ &    $-$ 0.88 &  1.02~ \\
	
	\multirow{1}{*}{\text{HNC}}   & $R_{e}$ (H-N) &  99.49~  & 0.57~  &  0.59  &  0.30~  & 1.00~  &  0.10   & 0.86~  \\
		& $R_{e}$ (C-N) & 116.88~ & 1.45~ & 1.35   & 0.77~ & 0.59~ &  $-$0.62   &  $-$0.05~ \\
	
	\multirow{1}{*}{\text{C$_{2}$H$_{2}$}}   & $R_{e}$ (H-C) & 106.17~ & 0.53~ &  0.46  &  0.32~ & 0.82~  &  0.21	 & $-$0.06~  \\ 
		& $R_{e}$ (C-C) & 120.36~ & 1.27~  & 0.90   &  0.49~ & 0.28~ &  $-$0.79   &  0.50~ \\
	
		\multirow{1}{*}{\text{H$_{2}$O}}   & $R_{e}$ (H-O) &  95.80~ &  0.77~ &  0.76  &  0.46~  & 1.08~  &  $-$0.06  & 0.96~   \\ 
	
		\multirow{1}{*}{\text{HNO}}   & $R_{e}$ (N-H) & 105.20~ & 1.68~ & 1.33  &  0.32~  &  2.90~ & 0.85   &   $-$0.59~   \\
		&$R_{e}$ (N-O)  & 120.86~  & 1.59~  &  1.58  & 1.04~  & $-$0.12~  &  $-$1.85   & $-$0.46~   \\ 
		
	\multirow{1}{*}{\text{HOF}}   & $R_{e}$ (O-H) & 96.87~  &  0.68~  &  0.72  &  0.30~  &  1.07~ &  $-$0.25   &  $-$0.55~  \\
		& $R_{e}$ (O-F) & 143.45~  & 4.18~ &  4.08  & 2.21~  &  1.02~  & $-$2.85    &   $-$1.52~ \\
		
	\multirow{1}{*}{\text{N$_{2}$H$_{2}$}}   & $R_{e}$ (H-N) & 102.88~  &  0.83~  & 0.79  &  0.41~ &  1.51~   &   0.25   & $-$0.18~   \\ 
		& $R_{e}$ (N-N) & 124.58~  & 1.66~  &  1.52 &  0.91~ &  0.17~   & $-$1.53    &  0.61~  \\
	
    \multirow{1}{*}{\text{CH$_{2}$O}}   & $R_{e}$ (C-O) & 120.47~  &  1.25~  & 1.12  &  0.10~  & 0.23~    &  $-$1.03   & 0.35~ \\
		& $R_{e}$ (C-H) & 110.07~ & 0.84~  & 1.05   & 0.66~  & 1.70~  &     0.67  &  $-$0.15~ 	\\
		
\multirow{1}{*}{\text{C$_{2}$H$_{4}$}}   & $R_{e}$ (H-C) & 108.07~  &  0.59~    &  0.54 & 0.33~ &  1.02~   &  0.29   & $-$0.12~   \\ 
	    & $R_{e}$ (C-C) & 133.07~ & 1.06~ & 0.99  &  0.61~ &  0.15~ &    $-$0.83   &  $-$0.11~  \\			
		
		\multirow{1}{*}{\text{NH$_{3}$}}   & $R_{e}$ (H-N) & 101.24~ &  0.74~  & 0.68 & 0.41~   & 0.92~  &  $-$0.05    &  $-$0.26~   \\ 
		
		\multirow{1}{*}{\text{CH$_{4}$}}   & $R_{e}$ (C-H) & 108.59~  &  0.70~ &  0.65 &  0.44~ & 0.94~ & 0.22     &   $-$0.18~ \\
		
		\multirow{1}{*}{\text{H$_{2}$S}}   & $R_{e}$ (H-S) & 133.56\footnotemark[2] &  0.96~ & 0.84  &  0.56~ &  1.67~  &  0.60  & $-$0.20~    	\\
       Ne$_2$ & $R_{e}$ & 310.00\footnotemark[3] & $-$7.55~   & $-$6.39 & 6.82~ & $-$1.19~   &  2.38   & 9.20~ \\
	  Ar$_2$ & $R_{e}$ & 375.80\footnotemark[3]  & 2.18~  & $-$1.49  &   7.86~ & 10.25~  &  27.15   & $-$0.40~  \\
	  Li$_{2}$ & $R_{e}$& 267.30\footnotemark[4] & 4.40~  &  4.10  & 3.11~  &  4.41~ &  4.76   &  7.30~   \\
      BF & $R_{e}$&126.69\footnotemark[4]   &  1.76~ & 1.72  &  0.62~  & 0.54~ &  $-$0.77  &   $-$0.25~   \\
      HCl & $R_{e}$& 127.46\footnotemark[4] & 0.85~ &  0.75 &  0.47~  & 1.49~  &  0.44  & $-$0.23~  \\
	  Cl$_{2}$ & $R_{e}$&  198.79\footnotemark[4] &  4.31~  & 3.13~	&  2.76~  & 2.49~  &  $-$0.20~  &  $-$0.26~ \\
	  $\text{NO}^{+}$& $R_{e}$ &106.56\footnotemark[4]  &  1.27~   & 1.18 &  0.52~   & 0.36~ & $-$1.25  &  1.30~   \\
      BeH$_{2}$ & $R_{e}$&  132.64\footnotemark[4]  &  0.60~ & 0.60  & 0.41~    & 1.03~ &  0.56~  &   0.06~  \\
		ME & $R_{e}$&  & 1.19~   &  1.01 & 1.16~ & 1.24~  &  0.61  & 0.50~ \\
		MAE &$R_{e}$&  &  1.63~   & 1.47  & 1.16~  & 1.31~  &  1.69  &  0.96~  \\
            MAPE &      &  & 1.12\%~  &  1.03\%~ & 0.67\%~ & 0.96\%~ & 0.85\%~ & 0.58\%~    \\
            [0.5ex]
		\multicolumn{9}{c}{Bond Angles (deg)} \\ [0.5ex]
	
	H$_2$O&  $\varTheta_{e}$  & {104.4776}~  & $-$1.3174~    &  $-$0.9752~ & $-$0.7107~ & $-${0.5571}~   & 0.2111~  &  $-${0.4648}~   \\
	
		HNO & $\varTheta_{e}$ & 108.2600~ & $-$0.1743~ & 0.0127~  & $-$0.2839~  & 0.3952~ &  0.5308~  &  $-$0.4711~   \\
		
		\multirow{1}{*}{\text{HOF}}   & $\varTheta_{e}$ & 97.2000~  & $-$0.8483~   & $-$0.7976~   & $-$0.0071~   & 0.6919~ &  1.8273~   & 0.7962~  \\
		
		N$_2$H$_2$ & $\varTheta_{e}$  & {106.3000}~ & $-$0.7775~  & $-$0.6138~   &  $-$0.4516~  &  $-${0.0890}~   &  0.5167~ &   $-${0.5530}~   \\
		
		CH$_2$O &$\varTheta_{e}$ (HCO) & {121.9330}~  &  0.0718~  & 0.1628~  &  $-$0.1907~ & {0.1857}~  &  0.0627~   &   $-${0.1500}~   \\
		&$\varTheta_{e}$ (HCH) & {116.1340}~  & $-$0.1425~  & $-$0.3241~  &  0.3812~  & $-${0.3714}~ &  $-$0.1254~   &  {0.3000}~   \\
		
		C$_2$H$_4$ & $\varTheta_{e}$ (HCC)  & {121.2000}~ &  0.2643~   &  0.2517~  & 0.2253~ & {0.5096} &  0.4529~   &  {0.1270}~   \\
		& $\varTheta_{e}$ (HCH)  & {117.6000} & $-${0.5285}~   &  $-$0.5037~ & $-${0.4498}~   &  $-${1.0193}~  & $-$0.9057~   &   $-${0.2550}~   \\
		
		NH$_3$ & $\varTheta_{e}$  & {106.6732} & $-$1.5073~ & $-$1.0967~ &  $-$0.7033~  & $-${0.8349}~  &  $-$0.0366~   &  $-${0.2090}~   \\
		
		CH$_4$&$\varTheta_{e}$  & {109.4710}~  & 0.0002~  & 0.0001~  &  0.0001~  &{0.0000}~ &      $-$0.0041~  & {0.0000}~   \\
		
		H$_2$S & $\varTheta_{e}$  & {92.1100}\footnotemark[2] &  0.1011~  &  0.1539~  & 0.0633~ & $-${0.3828}~  &  0.2117~   & {0.1270}~   \\

		ME & $\varTheta_{e}$  &  & $-$0.6570~ & $-$0.3391~   &  $-$0.2114~ & 0.1338~  &   0.2492~  &  $-$0.0684~ \\
		
		MAE & $\varTheta_{e}$& & {0.7880}~  &  0.4448~ & {0.3102}~  &  {0.4579}~ &  0.4441~  &  0.3139~ \\

         MAPE &      &  & ~0.49\%~  &    0.42\%~ & 0.29\%~ & 0.36\%~ &  0.42\%~ & 0.30\%~  \\
		
\end{longtable}

	
		

The MP2 results are taken from \onlinecite{nwchem,CCCBDB}, while all other results are computed using FHI-aims.  The frozen-core approximation is used in RPA and RPA+rSE calculations. Furthermore, the cc-pVQZ (QZ) basis set is used in PBE, PBE0 and MP2 calculations, except for rare-gas dimers where the aug-cc-pVQZ (aQZ) basis set is used. For the RPA+rSE calculations, the cc-pV5Z (5Z)  results (and aug-cc-pV5Z (a5Z) results for rare-gas dimers) are also provided to demonstrate the influence of basis size on RPA+rSE geometries across the molecular set. The basis dependence of RPA geometries has been checked in Ref.~\citenum{Muhammad_2022}, and here only the results of 5Z (a5Z) are shown. The reference geometries adopted here are mostly taken from \onlinecite{Filip/Poul/Jeppe/stanton/2002}.  These are regarded as \textit{empirical} equilibrium geometries, determined from a mixture of experimental rotational constants and theoretical vibration-rotation interaction constants obtained by the method of coupled cluster with single, double, and perturbative triple excitations (CCSD(T)) with the cc-pVQZ basis set.

 From Table~\ref{tab:comparison_of_results}, it is evident that the (RPA+rSE)@PBE approach consistently overestimates the bond lengths of small molecules with strong covalent bonds. Previously, we have shown that RPA itself systematically overestimates the bond lengths of small molecules, which is consistent with the general underbinding behavior of RPA. The rSE term correctly increases the bonding strength of molecules. However,  surprisingly, the bond length yielded by RPA+rSE does not change in the right direction as one may expect from the bonding strength increase. To understand the origin of this behavior, in Fig~\ref{fig:binding_CO-Ar2_a} we plotted the RPA@PBE and (RPA+rSE)@PBE binding energy curves of the CO molecule, which has a strong covalent bond. It can be seen that the magnitude of the rSE contribution increases with the bond length, which effectively shifts the equilibrium bond length to a larger value, as indicated by the arrows in the figure. Close inspection reveals that this is because the energy gap between the highest occupied molecular orbital (HOMO) and lowest unoccupied molecular orbital (LUMO) shrinks as the CO bond length increases. The rSE contribution strongly depends on the HOMO-LUMO gap in a roughly inverse way, as can be inferred from Eq.~\ref{eq:rSE_energy};  thus its magnitude gradually increases as the HOMO-LUMO gap reduces. This behavior is rather general for strongly bonded molecules. However, for the weak vdW bonding, adding the rSE term gives an opposite effect and the equilibrium bond length is correctly reduced, as demonstrated in Fig~\ref{fig:binding_CO-Ar2_b} for the Ar$_2$ dimer. As can be seen from the inset of Fig~\ref{fig:binding_CO-Ar2_b}, now the rSE contribution gradually decreases with the Ar$_2$ bond length.  The reason for this behavior is also simple: The HOMO-LUMO gap of Ar$_2$ does not decrease upon increasing 
 the distance between the two Ar atoms, because the constituent atom itself is closed-shelled and the inter-atomic interaction is of pure dispersion nature.   
 
\begin{figure}[ht!]
	\centering
       \subfigure[]{
        \includegraphics[width=0.38\textwidth]{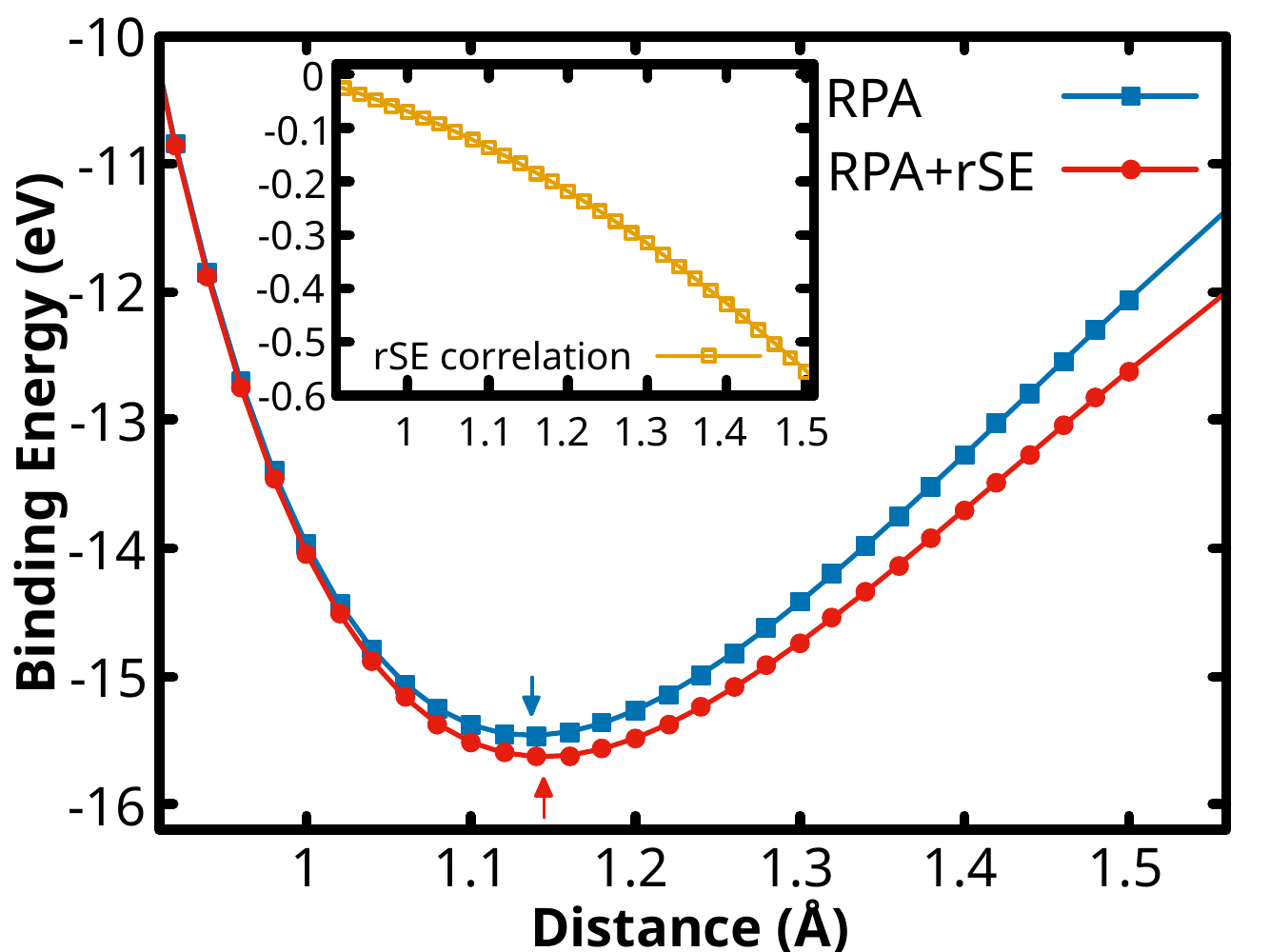}
        \label{fig:binding_CO-Ar2_a}
       }
       \subfigure[]{
		\includegraphics[width=0.38\textwidth]{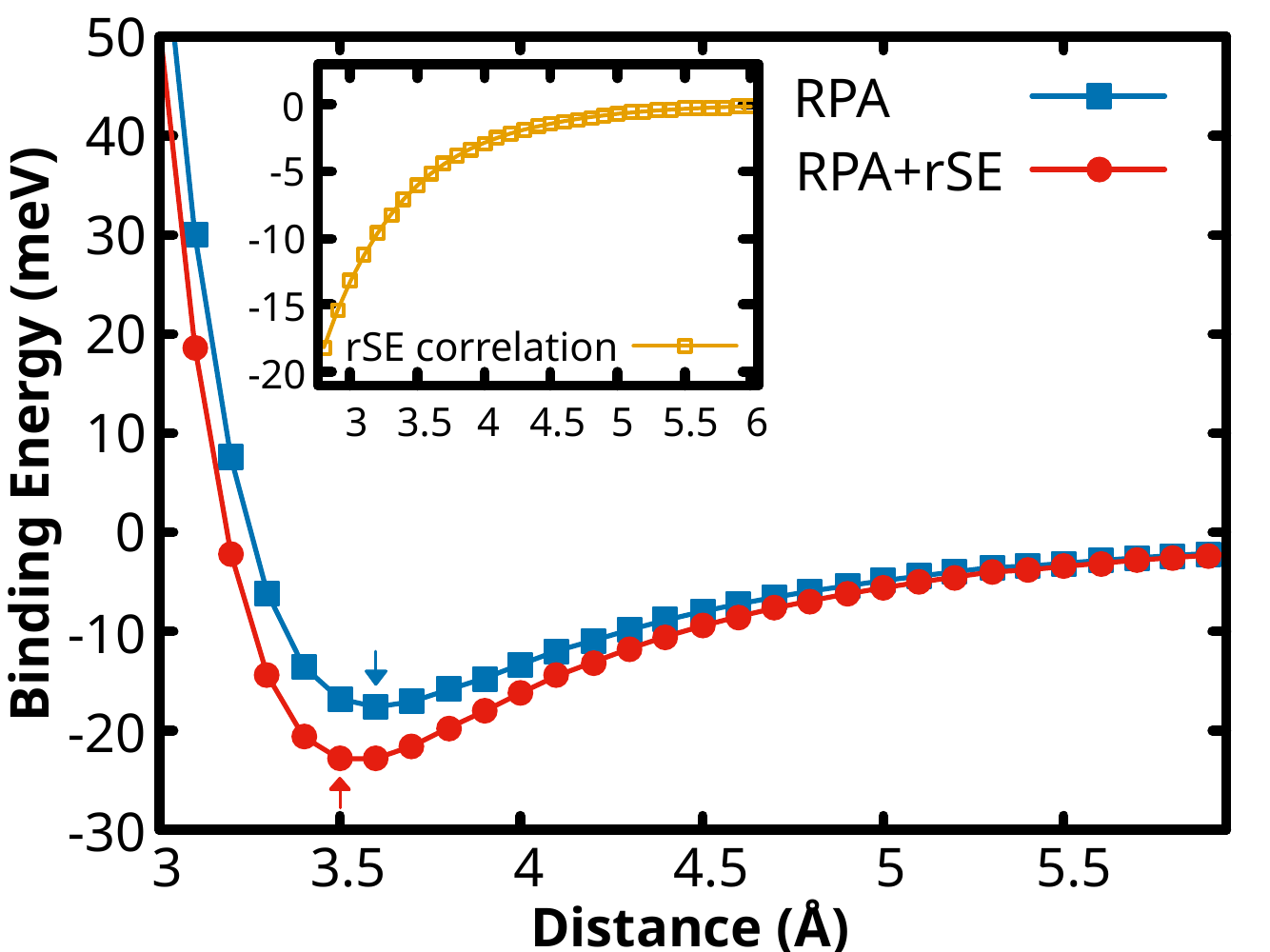} 
        \label{fig:binding_CO-Ar2_b}
        }
        \caption{\label{fig:binding_CO-Ar2}The RPA and RPA+rSE binding energy curves for CO (left panel) and Ar2 (right panel) molecules. The contributions from the rSE correction are shown in the inset.  The cc-pV5Z and aug-cc-pV5Z basis sets are used for CO and Ar$_2$ calculations, respectively. The basis set superposition error is not corrected. The equilibrium bond lengths determined by the two methods are indicated by arrows. }
\end{figure}

The error statistics presented in Table~\ref{tab:comparison_of_results} show that the mean error (ME) of the 
(RPA+rSE)@PBE bond lengths is comparable to that of RPA@PBE, while the mean absolute error (MAE) of the former is slightly larger. This is partly because the test set mainly consists of covalent molecules, for which the RPA error is relatively smaller. Regarding the polyatomic molecules, RPA+rSE tends to underestimate the bond angles and yields ME and MAE that are noticeably larger than those of RPA. Another important point is that, for both the bond lengths and bond angles, the RPA+rSE shows a pronounced basis set dependence; from QZ to 5Z (aQZ to a5Z), the ME and MAE of RPA+rSE are noticeably decreased.

\begin{table*} [ht!]
	\caption{\label{tab:comparison_of_results1} rPT2@PBE  bond lengths $R_e$ (in pm), in comparison with RPA@PBE and (RPA+rSE)@PBE results, obtained using  cc-pV5Z (aug-cc-pV5Z for rare-gas dimers) basis set.  
    The reference values are taken from the same sources as Table~\ref{tab:comparison_of_results}.}
\centering
	\scalebox{0.950}{
	\begin{tabular}{lcrrrr}
    \hline\hline 
		\multicolumn{1}{c}{\multirow{2}{*}{System}} &
		\multicolumn{1}{c}{\multirow{2}{*}{~~~~~~$R_e$~~~~~~~}} &
		\multicolumn{1}{c}{\multirow{2}{*}{~~~~Ref. \footnotemark[1] }} & \multicolumn{1}{c}{\multirow{2}{*}{~~~~~~~RPA/5Z}} & \multicolumn{1}{c}{\multirow{2}{*}{(RPA+rSE)/5Z}} &
		\multicolumn{1}{c}{\multirow{2}{*}{~~~~~~rPT2/5Z}} 
		 \\[2.0ex]
		\hline \hline 
		H$_2$ &  $R_{e}$ & 74.15 &    0.13 & 0.23 & $-$0.31  \\
		F$_2$ & $R_{e}$& 141.27 &  2.18 & 3.52 & $-$4.52   \\
		N$_2$ &  $R_{e}$& 109.76  & 0.68 & 1.12 & $-$0.88  \\
		HF &  $R_{e}$& 91.68 &  0.48 & 0.84 &  $-$0.52 \\
		BF & $R_{e}$&126.69\footnotemark[4] & 0.62 & 1.72 & $-$0.09  \\
		Cl$_2$  &   $R_{e}$&  198.79\footnotemark[4] &  2.76 & 3.13 & $-$0.90  \\
		HCl & $R_{e}$& 127.46\footnotemark[4]  &  0.47 & 0.75 &$-$0.28  \\
		CO & $R_{e}$ &   112.82 & 0.84 & 1.58 & $-$0.32 \\
		Li$_2$ & $R_{e}$& 267.30\footnotemark[4] & 3.11 & 4.10 & 7.34  \\ 
		Ne$_2$ & $R_{e}$ & 310.00\footnotemark[3] & 6.82 & $-$6.39 & $-$9.60  \\
		Ar$_2$ & $R_{e}$ & 375.80\footnotemark[3] & 7.86 & $-$1.49 & 1.86  \\ 
		ME & & & 2.36 & 1.55 & $-$0.75 \\
		MAE  & & & 2.36 & 2.26 & 2.42 \\
		\hline \hline 
		$^a$Ref.~\onlinecite{Filip/Poul/Jeppe/stanton/2002} & $^b$Ref.~\onlinecite{COOK/1975}
		& $^c$Ref.~\onlinecite{Tang/Toennies:2003} & $^d$Ref.~\onlinecite{CCCBDB} 
 		\end{tabular}}
		\end{table*}

For the conventional approaches, PBE shows a well-known overestimation behavior for the bond lengths, similar to RPA and RPA+rSE. On the other hand, PBE0 and MP2 show a more balanced performance for the strongly bonded molecules, resulting in an overall smaller ME for the bond lengths (and also for the bond angles in the case of MP2). The distinction between MP2 and RPA is that MP2 has both direct and exchange terms up to the second order whereas RPA (here the \textit{direct} RPA) only has direct ring terms up to infinite order.  Based on these results, we conjecture that the systematic overestimation of bond lengths by RPA and RPA+rSE might be due to the lack of higher-order exchange-type terms in these approaches. To check this conjecture, we chose a set of diatomic molecules and used the renormalized  second-order perturbation theory (rPT2) \cite{Ren/etal:2013} to determine their equilibrium bond lengths. Compared to RPA+rSE, rPT2 also contains the second-order screened exchange (SOSEX) term \cite{Grueneis/etal:2009,Paier_2012}, which represents a summation of a special type of exchange-type diagrams to infinite order. Since we don't have the analytical gradients for the SOSEX term yet, the rPT2 bond lengths for the diatomic molecules are directly determined from the equilibrium point of the binding energy curve.  In Table~\ref{tab:comparison_of_results1}, we presented the calculated rPT2 bond lengths for a subset of 11 molecules, in comparison with the values obtained by RPA and RPA+rSE, all based on the PBE reference. It can be seen that the rPT2 indeed reduces the bond lengths for most of the molecules, shifting the ME from a positive value of 1.55 pm to a negative value of $-0.75$ pm. This confirms the conjecture that the SOSEX term has an opposite effect as rSE for determining the bond length of strongly bonded molecules.

\subsection{RPA+rSE geometries and energy ranking for low-Energy isomers of (H$_2$O)$_6$ clusters}
\label{sec:water_hexamer}
The water hexamers, i.e., water clusters with six H$_2$O molecules, can arrange themselves into several energetically very close isomer structures \cite{Hincapie/etal:2010}. In particular, the four isomers with the lowest energies, termed ``prism", ``cage", ``book" and ``cyclic", exhibit total energy differences of no more than 10-20 meV/H$_2$O.
As such, the energy ranking of these four isomers has become a standard test case for benchmarking the accuracy of quantum chemical electronic structure methods. High-level quantum chemistry methods like MP2 \cite{Biswajit/etal:2008} and CCSD(T) \cite{Olson/etal:2007}, along with diffusion Monte Carlo \cite{Biswajit/etal:2008}, have ascertained the ground-state energy rankings of these isomers (neglecting zero-point energy and temperature effects). Namely, these calculations consistently show that the prism isomer is the most stable, followed by the ``cage", ``book", and ``cyclic" isomers. However, many semi-local and hybrid density functional approximations (DFAs) fail to predict the most stable isomer or the correct energy hierarchy of these four isomers \cite{Biswajit/etal:2008}.

Previously, we have checked the performance of the RPA method for this test set \cite{Muhammad_2022}. Based on the RPA-optimized geometries, RPA@PBE yields the correct energy ordering of these four isomers. In Ref.~\citenum{Muhammad_2022}, we also performed single-point RPA+rSE calculations on top of RPA geometries, where not only the energy ordering, but also the quantitative energy differences between these four isomers, are very close to the results of CCSD(T). 

The question is, what happens if one uses consistently the RPA+rSE geometries for the RPA+rSE energy calculations. Now, with the analytical RPA+rSE gradients, this question can be addressed. In Table~\ref{tab:dissociation_ene_water_hexamer}, we present the (RPA+rSE)@PBE binding energies for the four (H$_2$O)$_6$ isomers, calculated on top of both RPA and RPA+rSE geometries. While the geometries are optimized using the cc-pVQZ basis set, the energies are calculated with several different GTO basis sets \cite{Dunning:1989}, as well as the NAO-VCC-4Z \cite{IgorZhang/etal:2013} basis set, which belongs to the ``correlation consistent'' NAO basis sets designed for correlated calculations.  Additionally, we mention that for numerically accurate (RPA+rSE)@PBE single point energies, we included extra $5g$ hydrogen-like functions (with an effective charge $Z=6$) to create additional ABFs (denoted by the \textit{"for\_aux"} tag in FHI-aims) \cite{Ihrig/etal:2015, Ren/etal:2021} to mitigate the numerical errors associated with LRI. In this way, the RPA energetics obtained within the LRI scheme is comparable to the traditional RI-RPA method with the global Coulomb metric.

From Table~\ref{tab:dissociation_ene_water_hexamer}, one can see that, except for the cc-pVQZ basis set, the binding energies obtained with larger GTO basis sets, as well as NAO-VCC-4Z, are fairly close to one another. In particular, it seems the $4Z$-level NAO basis set can yield RPA+rSE results of comparable quality as the $5Z$ GTO basis set. As for the energy differences between different isomers, the discrepancies are mostly within 1 meV/H$_2$O.  The situation does not change when one goes to the complete basis set (CBS) limit extrapolated from aQZ and a5Z results \cite{Helgaker/etal:1997}. The geometry effect is also marginal in the present case. Using the RPA+rSE geometries instead of the RPA ones, there are no significant changes in the binding energies, and their differences also agree with their counterparts obtained with RPA geometries within 1 meV/H$_2$O.

\begin{table*} [ht!]
	\caption{\label{tab:dissociation_ene_water_hexamer}  The (RPA+rSE)@PBE dissociation energies (in meV/H$_{2}$O) 
	of the four water hexamers obtained using various basis sets, with structure relaxed 
	using RPA@PBE and (RPA+rSE)@PBE geometries. The dissociation energies for the lowest-energy (prism) isomer are highlighted in bold, and the energy difference between other isomers and the lowest-energy one are indicated in parenthesis. The geometries are relaxed using the cc-pVQZ basis set. An additional $5g (Z=6)$ hydrogen-like ``for\_aux" basis function \cite{Ihrig/etal:2015} is employed to create the additional ABFs to enhance the accuracy of LRI \cite{Ihrig/etal:2015}. The frozen-core approximation is used in all calculations.}

	\begin{tabular}{lcccl} 
		\hline\hline 
		\multicolumn{1}{c}{\multirow{2}{*}{Basis Set}} &
		\multicolumn{1}{c}{\multirow{2}{*}{Prism}} &
		\multicolumn{1}{c}{\multirow{2}{*}{Cage}} & \multicolumn{1}{c}{\multirow{2}{*}{Book}} & \multicolumn{1}{c}{\multirow{2}{*}{Cyclic}} 
		\\     
		& & & &  \\ 
		\hline 	\\ [-0.5ex]
        \multicolumn{5}{c}{Structure Relaxation}\\
		\multicolumn{5}{c}{RPA@PBE/cc-pVQZ}  \\ [3ex]
        cc-pVQZ & \bf{384.3} & 380.9(0.4)  &  371.3(13.0) & 355.1(29.20)    \\
 		cc-pV5Z & \bf{358.0} & 355.7(2.3)  &  349.3(8.7) & 337.4(20.6)    \\
 		NAO-VCC-4Z & \bf{357.3} & 355.4(1.9)  &  349.0(8.3) & 336.3(21.0)    \\
 		aug-cc-pVQZ & \bf{358.8} & 356.4(2.4)  &  349.6(9.2) & 337.1(21.7)    \\
 		aug-cc-pV5Z & \bf{352.8} & 350.3(2.5)  &  343.9(8.9) & 331.5(21.3)    \\
 	    CBS(aQZ,a5Z) & \bf{346.4} & 344.0(2.4)  &  337.8(8.6) & 325.7(20.7)    \\ [3ex]
        
		\multicolumn{5}{c}{Structure Relaxation}\\
		\multicolumn{5}{c}{(RPA+rSE)@PBE/cc-pVQZ}  \\ [3ex]
	
 		cc-pVQZ & \bf{386.0} & 382.8(3.2)  &  373.4(12.6) & 357.3(28.7)    \\
 		cc-pV5Z & \bf{358.4} & 355.9(2.5)  & 349.6(8.8)  & 337.8(20.6) \\
 		NAO-VCC-4Z & \bf{358.2}   & 356.0(2.2)  &  349.7(8.5)  & 337.1(21.1) \\
 		aug-cc-pVQZ & \bf{359.0}  & 356.4(2.6) & 349.7(9.3)  & 337.3(21.7) \\
 		aug-cc-pV5Z & \bf{351.9}  & 349.1(2.8) & 342.9(9.0) &  331.1(20.8)   \\
 	    CBS(aQZ,a5Z) &  \bf{344.5} & 341.4(3.1) & 335.8(8.7) & 324.6(19.9) \\ 
		\hline\hline
	\end{tabular}
\end{table*}

 \subsection{WATER27 Dataset}
 \label{sec:water27}

      The WATER27 dataset is part of the GMTKN24\cite{Grimme_2010} and GMTKN30\cite{Grimme_2011} benchmark suites. It consists of 27 water clusters introduced by Bryantsev \textit{et al.}\cite{Bryantsev_2009,Anacker/Friedrich:2014} where reference dissociation energies of neutral, protonated, and deprotonated water clusters are provided. 
    The size of these clusters ranges from dimers ($n=2$) to dodecamers ($n=20$).
      Originally, the reference energies are obtained by MP2/CBS with higher-order corrections from CCSD(T)/aug-cc-pVDZ \cite{Bryantsev_2009}. 
      In the present work, the reference CCSD(T)/CBS values are taken from the work of Manna \textit{et al}.\cite{Manna_2017}, which provided the most accurate dissociation energies at the CCSD(T) level to date.

In this work, we calculate the dissociation energies for the WATER27 test set by employing the RPA and RPA+rSE methods. The RPA and RPA+rSE dissociation energies are calculated with finite GTO basis sets and extrapolated to the CBS limit. 
The CBS limit for RPA and RPA+rSE is obtained by independently extrapolating the exact exchange energy and the RPA/RPA+rSE correlation energy. To determine the CBS limit of the exact-exchange energy, the approach proposed by Zhong \textit{et al.}\cite{Zhong_2008} was applied, 
\begin{equation*}
    E_{x}^{EX}(X,Y)=\dfrac{e^{-a\sqrt{Y}}E_{x}^{EX}(X)-e^{-a\sqrt{X}}E_{x}^{EX}(Y)}{e^{-a\sqrt{Y}}-e^{-a\sqrt{X}}}
\end{equation*}
where $X$ and $Y$ represent the cardinal number in the Dunning basis set, and the constant $a$ is set to the suggested value of $6.30$.
For the RPA correlation energy, a conventional two-point extrapolation procedure proposed by Halkier \textit{et al.}\cite{Halkier_1998} is used.

 Previously, Chedid \textit{et al}.\cite{Chedid/Jocelyn/Eshuis:2021}  calculated the RPA dissociation energies for the WATER27 dataset in the CBS limit, finding that RPA underbinds these water clusters by about 7 kcal/mol on average, performing even worse than, e.g., the B3LYP functional. Subsequently, Tahir \textit{et al}.\cite{Muhammad_2024} performed the RPA+rSE calculations for the WATER27 dataset. It was found that at finite GTO basis set, RPA+rSE shows an opposite overbinding behavior for these water clusters. However, the amount of overbindings reduces as the basis size increases, and one eventually ends up with a rather small MAE for RPA+rSE at the CBS limit. This holds for RPA+rSE calculations based on both the original B3LYP geometries and the RPA geometries. In this work, with the RPA+rSE geometries available, we further check what happens if the calculations are done using the RPA+rSE geometries. To this end, we performed RPA@PBE and (RPA+rSE)@PBE calculations based on three types of geometries, i.e., the original geometries of the WATER27 dataset obtained using the B3LYP functional, the RPA geometries and the RPA+rSE geometries. For the latter two types, the geometry relaxation was carried out using both cc-pVQZ and NAO-VCC-4Z basis sets, resulting in two sets of geometries of each type. For the energy calculations, both cc-pV5Z and cc-pV6Z basis sets are used. In Table~\ref{mean_deviation_WATER27},  we present the error analysis of RPA@PBE and (RPA+rSE)@PBE results obtained using  5Z and 6Z basis sets, as well as CBS(5Z,6Z) based on different sets of geometries. For comparison, the results from three doubly hybrid functionals, taken from ~\onlinecite{Chedid/Jocelyn/Eshuis:2021}, are shown in the top part of Table~~\ref{mean_deviation_WATER27}, to represent the current state of the art. The actual dissociation energies of each individual system in the WATER27 test set, obtained at the CBS limit, are presented in the Supporting Information (SI). The RPA+rSE optimized geometries using cc-pVQZ basis sets are also provided in SI. 

 In Table~\ref{mean_deviation_WATER27}, a negative ME means underbinding and a positive ME means overbinding. Consistent with previous findings \cite{Chedid/Jocelyn/Eshuis:2021,Muhammad_2024}, RPA significantly underbinds water clusters, while RPA+rSE shows an opposite behavior. For all geometries, the magnitude of the RPA underbinding increases with the basis size, while the amount of  overbinding for RPA+rSE gets reduced as one approaches the CBS limit. At the CBS limit, our RPA calculation yields a 7.64 kcal/mol overbinding, consistent with the 7.40 kcal/mol reported in Ref.~\citenum{Chedid/Jocelyn/Eshuis:2021}. This value increases by approximately 1 kcal/mol when the calculations are done with the RPA geometries, and by 2 kcal/mol if the RPA+rSE geometries are used. As for RPA+rSE, the amount of the overbinding is about 1.41 kcal/mol at the original B3LYP geometries,  and it reduces to about 1 kcal/mol with the RPA geometries and to about 0.9 kcal/mol with RPA+rSE geometries, if the cc-pVQZ basis set is used in the geometry relaxation. Furthermore, if the RPA and RPA+rSE geometries are optimized with the NAO-VCC-4Z basis set, the corresponding magnitude of overbinding, as measured by ME, is further reduced by about 0.1 eV.

Such a detailed study is merely served to examine the dependence of RPA and RPA+rSE energetics on detailed geometries of the water clusters. Here we see that the dependence is also rather mild. For example, although the ME and MAE of RPA+rSE dissociation energies are reduced when one goes from the B3LYP geometries to the RPA geometries, and to the RPA+rSE geometries, we shouldn't put too much emphasis on such slight improvement, as the original CCSD(T) reference results are obtained based on the B3LYP geometries.

\begin{table*} [ht!] 
    	\caption{ \label{mean_deviation_WATER27} Mean deviations (MDs), mean absolute deviations (MADs), and maximum absolute deviations (Max) in kcal/mol of RPA@PBE and (RPA+rSE)@PBE dissociation energies for the WATER27 test set \cite{Bryantsev_2009} obtained using different basis sets and geometries, with respect to  the reference energies taken from  Ref.~\citenum{Manna_2017}. The negative and positive MD mean underbinding and overbinding, respectively. }
     \scalebox{.85}{
    	\begin{tabular}{lrrr}
    		\hline\hline 	
    	\multicolumn{1}{c}{\multirow{2}{*}{Method}} &
    		\multicolumn{1}{c}{\multirow{2}{*}{MD}} & \multicolumn{1}{c}{\multirow{2}{*}{~~~~~~~~MAD}} &  \multicolumn{1}{c}{\multirow{2}{*}{~~~~~~~~~~MAX}}  \\\\
      \hline
      RPA@PBE [QZVPP]$^{\textcolor{black}{\ast}}$ & $-$2.90 &  4.90  & 16.40 \\
       RPA@PBE [CBS(5Z,6Z)]$^{\textcolor{black}{\ast}}$ & $-$7.40 & 7.40 & 25.90 \\
       M06-2X-D3(0)       &  3.40  &  3.70   &  10.00\\
       $\omega$B97X-D3(0) &  2.20 & 2.30  & 7.40\\
       DSD-BLYP-D3(0)     &  1.20 & 1.30  & 5.50 \\ [1.ex]
       \multicolumn{4}{c}{structure relaxation B3LYP/6-311++G(2d,2p)}\\
        [1.ex]
         RPA@PBE [QZVPP]$^{\textcolor{black}{\dagger}}$ & $-$2.74 &  5.02  & 16.34 \\
          (RPA+rSE)@PBE [QZVPP]$^{\textcolor{black}{\dagger}}$ & 7.89 & 7.89 & 18.19\\
        RPA@PBE [5Z]$^{\textcolor{black}{\dagger}}$ & $-$3.78 & 5.31 & 18.77 \\
        RPA@PBE [6Z]$^{\textcolor{black}{\dagger}}$ & $-$5.52 &  5.74 & 21.54\\
        (RPA+rSE)@PBE [5Z]$^{\textcolor{black}{\dagger}}$ &  6.76 &  6.76 & 15.35\\
        (RPA+rSE)@PBE [6Z]$^{\textcolor{black}{\dagger}}$ &  4.40 &  4.40 & 10.86\\
         RPA@PBE [CBS(5Z,6Z)]$^{\textcolor{black}{\dagger}}$ & $-$7.64 & 7.64 & 25.82 \\
       (RPA+rSE)@PBE [CBS(5Z,6Z)]$^{\textcolor{black}{\dagger}}$ & 1.40  & 1.41  & 4.22\\
        [1.ex]
       \multicolumn{4}{c}{structure relaxation RPA@PBE/cc-pVQZ}\\
        [1.ex]
        RPA@PBE [5Z]$\ddagger$ & $-$4.17 & 5.59 & 21.45  \\  
        RPA@PBE [6Z]$\ddagger$ & $-$5.99 & 6.16 & 25.55  \\
        RPA@PBE [CBS(QZ,5Z)]$\ddagger$ & $-$7.76 & 7.89 & 30.34 \\ 
        RPA@PBE [CBS(5Z,6Z)]$\ddagger$ & $-$8.34 & 8.34 &  31.59 \\
        (RPA+rSE)@PBE [5Z]$\ddagger$ & 6.63 & 6.63 &  15.81 \\ 
        (RPA+rSE)@PBE [6Z]$\ddagger$ & 4.16 & 4.16 &  10.20 \\ 
        (RPA+rSE)@PBE [CBS(QZ,5Z)]$\ddagger$ &  1.67 & 1.72 & 5.44 \\
        (RPA+rSE)@PBE [CBS(5Z,6Z)]$\ddagger$  & 0.92 & 1.08 & 3.71 \\ [1.ex]
       \multicolumn{4}{c}{structure relaxation (RPA+rSE)@PBE/cc-pVQZ}\\
        [1.ex]
        RPA@PBE [5Z]$\ddagger$ &  $-$5.22 & 6.33 & 25.84 \\
        RPA@PBE [6Z]$\ddagger$ & $-$7.05 & 7.21 & 28.85 \\  
        RPA@PBE [CBS(QZ,5Z)]$\ddagger$ &  $-$9.05 & 9.17 & 35.19  \\ 
        RPA@PBE [CBS(5Z,6Z)]$\ddagger$ & $-$9.32 & 9.32 & 33.46\\
         (RPA+rSE)@PBE [5Z]$\ddagger$ & 6.42 & 6.42 & 15.76 \\
         (RPA+rSE)@PBE [6Z]$\ddagger$ &  3.93   &  3.93   &  8.88  \\
        (RPA+rSE)@PBE [CBS(QZ,5Z)]$\ddagger$ &  1.13 & 1.48 & 5.23 \\
        (RPA+rSE)@PBE [CBS(5Z,6Z)]$\ddagger$  &  0.76 & 0.82 & 3.55 \\
        \multicolumn{4}{c}{structure relaxation RPA@PBE/NAO-VCC-4Z}\\
        [1.ex]
         RPA@PBE [5Z]$\ddagger$ &  $-$4.50 & 5.85 & 22.64  \\
         RPA@PBE [6Z]$\ddagger$ &  $-$6.30 & 6.48 & 25.64 \\
         RPA@PBE [CBS(5Z,6Z)]$\ddagger$ & $-$8.57 & 8.57 & 29.41\\ 
         (RPA+rSE)@PBE [5Z]$\ddagger$ &  6.46 & 6.46 &  13.55  \\
        (RPA+rSE)@PBE [6Z]$\ddagger$ &   4.01 & 4.01 &  9.87\\
         (RPA+rSE)@PBE [CBS(5Z,6Z)]$\ddagger$  & 0.86 & 0.91 & 3.39 \\
        \multicolumn{4}{c}{structure relaxation (RPA+rSE)@PBE/NAO-VCC-4Z}\\
        [1.ex]
         RPA@PBE [5Z]$\ddagger$ &  $-$5.46 & 6.57 & 26.53 \\
         RPA@PBE [6Z]$\ddagger$ &  $-$7.30 & 7.30 & 29.61 \\
         RPA@PBE [CBS(5Z,6Z)]$\ddagger$ & $-$9.59 & 9.59 & 34.41 \\ 
         (RPA+rSE)@PBE [5Z]$\ddagger$ &  6.26   &  6.26   &  12.95  \\
        (RPA+rSE)@PBE [6Z]$\ddagger$ &  3.76   &  3.76   &  9.77  \\
         (RPA+rSE)@PBE [CBS(5Z,6Z)]$\ddagger$  &  0.58 & 0.70 & 3.18 \\
     		\hline\hline 
      \multicolumn{4}{l}{$^{\textcolor{black}{\ast}}$Ref~\citenum{Chedid/Jocelyn/Eshuis:2021} ~~~$^{\textcolor{black}{\dagger}}$Ref~\citenum{Muhammad_2024} ~~~$\ddagger$This work (results obtained using FHI-aims)}\\
    	\end{tabular}
      }
    \end{table*}

   \section{Conclusion} 
   \label{sec:conclusion}
    In this work, we have derived and implemented the theoretical formalism for analytical RPA+rSE force calculations within the atomic-orbital basis set framework. Benchmark tests against the numerical forces obtained by the FD method proved the correctness of our RPA+rSE analytical force implementation. This enabled us to relax molecular geometries at the RPA+rSE level. Surprisingly, we found that (RPA+rSE)@PBE overestimates the bond lengths of the small covalently bonded molecules to an even larger extent than the standard RPA@PBE scheme. In contrast, the bond lengths of weakly bonded molecules get reduced. This is due to the different behaviors of the HOMO-LUMO gap as a function of the bond length for these two types of molecules. We then apply RPA+rSE to relax the structures of the water clusters. For the water hexamers, the correct energy hierarchy of the low-lying isomers is retained, as expected, by replacing the RPA geometries with the RPA+rSE ones, and even the dissociation energies of the clusters undergo very little changes ($\sim$ 1 meV/atom). For the larger WATER27 test set, the dissociation energies yielded by RPA+rSE are one order of magnitude more accurate than the RPA ones at the CBS limit for both RPA and RPA+rSE geometries.  Using the RPA+rSE geometries instead of the RPA ones results in a roughly 10\% increase in the ME/MAE for RPA and a 20\% decrease for RPA+rSE.

    In brief, our work opens the possibility of optimizing molecular geometries at the level of RPA+rSE. However, for the limited test calculations we have performed so far, we don’t observe an obvious advantage of the (RPA+rSE)-optimized geometries over the RPA ones. More investigations are needed to have a full picture of the influence on the geometries upon including the rSE corrections. To this end, more reliable references of the molecular structures from experiments or more accurate methodologies such as CCSD(T) are also in need.

    \section*{Acknowledgments}
  We acknowledge the funding support by the Strategic Priority Research Program of Chinese Academy of Sciences under Grant No. XDB0500201 and by the National Key Research and Development Program of China (Grant Nos. 2022YFA1403800 and 2023YFA1507004).  This work was also supported by the Chinese National Science Foundation Grant Nos 12134012, 12374067, 12188101, 12274254, and T2222026. The numerical calculations in this study were partly carried out on the ORISE Supercomputer.  

\begin{suppinfo}
The following file is available free of charge.
\begin{itemize}
  \item supporting\_info.pdf:
  The file contains the following items:
  \begin{enumerate}
      \item CBS(5Z,6Z) RPA@PBE and (RPA+rSE)@PBE binding energies of WATER27 dataset obtained using structures relaxed with cc-pVQZ basis sets;
       \item CBS(5Z,6Z) RPA@PBE and (RPA+rSE)@PBE binding energies of WATER27 dataset obtained using structures relaxed with cc-pV5Z basis sets;
        \item CBS(5Z,6Z) RPA@PBE and (RPA+rSE)@PBE binding energies of WATER27 dataset obtained using structures relaxed with NAO-VCC-4Z basis sets; 
        \item All-electron (RPA+rSE)@PBE optimized atomic coordinates of WATER27 dataset using cc-pVQZ basis sets.
  \end{enumerate}
  \end{itemize}
\end{suppinfo}
    
\begin{appendix}
\section{Further derivations}
\subsection{Derivation of $\dfrac{df_{pq}}{dc_{i,s}}$ (Eq.~\ref{eq:dfbydc}) } \label{sec:df-dc}
The derivative of $f_{pq}$ defined in Eq.~(\ref{eq:fock_matrix}) with respect to the KS eigenvectors is given as 
\begin{equation}
f_{pq,is}^{(1)} = \frac{d f_{pq}}{d c_{is}} = \frac{d}{d c_{is}} \left(-\sum_{\mu} \sum_{m}^{occ} O_{pm}^{\mu} O_{mq}^{\mu} -V^\text{xc}_{pq} + \delta_{pq}\epsilon_{p} \right)\,,
\end{equation}
Using Eq.~(\ref{eq:O_integrals}), we have 
\begin{equation}
\begin{split}
\dfrac{df_{pq}}{dc_{is}}=&- \sum_{\mu}\sum_{m}^{occ} \sum_{jk} \left( \frac{d c_{jp}}{d c_{is}} c_{km} + c_{jp} \frac{d c_{km}}{d c_{is}} \right) Q_{jk}^{\mu} O_{mq}^{\mu}  \\& 
-\sum_{\mu}\sum_{m}^{occ} \sum_{jk} \left( \frac{d c_{jm}}{d c_{is}} c_{kq} + c_{jm} \frac{d c_{kq}}{d c_{is}} \right) Q_{jk}^{\mu} O_{pm}^{\mu}  \\&  
-\sum_{jk}\bigg( \dfrac{dc_{jp}}{dc_{is}} c_{kq} +c_{jp}\dfrac{dc_{kq}}{dc_{is}} \bigg) V^\text{xc}_{jk}
\end{split}
\end{equation}
Using $\dfrac{dc_{jp}}{dc_{is}}=\delta_{ij}\delta_{ps}$, the above equation can be simplified as
\begin{equation}
\begin{split}
\dfrac{df_{pq}}{dc_{is}}=&  -\sum_{\mu}\sum_{m}^{occ} \sum_{jk} \left( \delta_{ij} \delta_{ps} c_{km} + c_{jp} \delta_{ik} \delta_{ms} \right) Q_{jk}^{\mu} O_{mq}^{\mu}  \\& 
- \sum_{\mu}\sum_{m}^{occ} \sum_{jk} \left( \delta_{ij} \delta_{ms} c_{kq} + c_{jm} \delta_{ik} \delta_{qs} \right) Q_{jk}^{\mu} O_{pm}^{\mu} \\& 
- \sum_{jk} \bigg ( \delta_{ij}\delta_{ps}c_{kq}+c_{jp} \delta_{ik}\delta_{qs} \bigg)V^\text{xc}_{jk} \\
= &  -\sum_{\mu}\sum_{m}^{occ} \left( \sum_{k} \delta_{ps} c_{km} Q_{ik}^{\mu} + \sum_{j} c_{jp} \delta_{ms}  Q_{ji}^{\mu}\right) O_{mq}^{\mu}  \\& 
- \sum_{\mu}\sum_{m}^{occ}  \left(\sum_{k}  \delta_{ms} c_{kq} Q_{ik}^{\mu} + \sum_{j}c_{jm} \delta_{qs} Q_{ji}^{\mu} \right)  O_{pm}^{\mu} \\& 
- \sum_{k}\delta_{ps}c_{kq}V^\text{xc}_{ik}-\sum_{j}c_{jp}\delta_{qs} V^\text{xc}_{ji}\,.
\end{split}
\end{equation}
Rearranging the above equation, we finally arrive at
\begin{equation}
\begin{split}
    \dfrac{df_{pq}}{dc_{is}} =&-\sum_{\mu}\sum_{m}^{occ}\sum_{j}  \left( \delta_{ps} c_{jm} Q_{ij}^{\mu} + c_{jp} \delta_{ms}  Q_{ji}^{\mu}\right) O_{mq}^{\mu}  \\& 
- \sum_{\mu}\sum_{m}^{occ} \sum_{j} \left( \delta_{ms} c_{jq} Q_{ij}^{\mu} + c_{jm} \delta_{qs} Q_{ji}^{\mu} \right)  O_{pm}^{\mu} \\& 
- \sum_{j}\left(\delta_{ps}c_{jq}V^\text{xc}_{ij}+c_{jp}\delta_{qs} V^\text{xc}_{ji} \right)\,.
\end{split}
\end{equation}

\subsection{ Derivation of $\dfrac{df_{pq}}{dV_{\mu \mu'}}$ (Eq.~\ref{eq:dfbydV})} \label{sec:df-dv}
The derivative of $f_{pq}$ defined in Eq.~(\ref{eq:fock_matrix}) with respect to $V_{\mu\mu'}$ is given by
\begin{equation}
    \begin{split}
\dfrac{df_{pq}}{dV_{\mu \mu'}}&=
    -\sum_{m}^{occ} \sum_{\nu} \dfrac{d}{dV_{\mu \mu'}}  \Big ( O^{\nu}_{pm} O^{\nu}_{mq} \Big) \, .
    \end{split}
\end{equation}
Using Eqs.~\ref{eq:O_integrals} and \ref{eq:Q_integrals}, we have
\begin{equation}
    \begin{split}
\dfrac{df_{pq}}{dV_{\mu \mu'}} =&-\sum_{m}^{occ}\sum_{\nu \nu'} \sum_{ij}   \Bigg [  c_{ip}\dfrac{dV^{\frac{1}{2}}_{\nu \nu'}}{dV_{\mu \mu'}}C^{\nu'}_{ij} c_{jm} O^{\nu}_{mq} 
    +O^{\nu}_{pm} c_{im}c_{jq} \dfrac{dV^{\frac{1}{2}}_{\nu \nu'}}{dV_{\mu \mu'}}C^{\nu'}_{ij} \Bigg] \\
    = & -\dfrac{1}{2}\sum_{m}^{occ}\sum_{ij} \sum_{\nu\nu'\nu''} \Bigg [  c_{ip}V^{-\frac{1}{2}}_{\nu \nu''} \dfrac{dV_{\nu' \nu''}}{dV_{\mu \mu'}}C^{\nu'}_{ij} c_{jm} O^{\nu}_{mq} +O^{\nu}_{pm} c_{im}c_{jq}  V^{-\frac{1}{2}}_{\nu \nu''} \dfrac{dV_{\nu' \nu''}}{dV_{\mu\mu'}}C^{\nu'}_{ij}\Bigg]\, .
    \end{split}
\end{equation}
Since $\dfrac{dV_{\nu' \nu''}}{dV_{\mu \mu'}}=\delta_{\mu\nu'}\delta_{\mu'\nu''}$, the above equation is simplified to 
\begin{equation}
    \begin{split}
&\dfrac{df_{pq}}{dV_{\mu \mu'}}=-\dfrac{1}{2}\sum_{m}^{occ}\sum_{ij} \sum_{\nu}\Bigg [ c_{ip}C^{\mu}_{ij} c_{jm} O^{\nu}_{mq}V^{-\frac{1}{2}}_{\nu \mu'}+ c_{im}C^{\mu}_{ij}c_{jq} O^{\nu}_{pm}V^{-\frac{1}{2}}_{\nu \mu'}\Bigg]
    \end{split}
\end{equation}

Swapping the indices $q$ and $q$ in the second term, one can see that the two terms become identical. Further
noticing $\sum_{ij}c_{ip}C^{\mu}_{ij}c_{jm}=\sum_{\nu''}V^{-\frac{1}{2}}_{\mu \nu''} O^{\nu''}_{pm}$, we finally
arrive at
\begin{equation}
    \begin{split}
\dfrac{df_{pq}}{dV_{\mu \mu'}}&=-\sum_{m}^{occ} \sum_{\nu \nu''}V^{-\frac{1}{2}}_{\mu \nu''} O^{\nu''}_{pm}  O^{\nu}_{mq}V^{-\frac{1}{2}}_{\nu \mu'}\,.
    \end{split}
\end{equation}
\end{appendix}

\bibliography{./CommonBib}

\IfFileExists{SupportingInfo.pdf}{
	

\section*{Supplementary material}

In Table S3, the binding energies of RPA@PBE, (RPA + rSE) are presented with Gaussian basis sets CBS (5Z, 6Z) with the structure relaxed with the given method using NAO-VCC-4Z basis sets. For $(H_2O)_{20}$, the structures are relaxed using the smaller NAO-VCC-3Z basis set. The PBE binding energies were calculated using the NAO-VCC-4Z basis set with PBE/NAO-VCC-4Z relaxed geometries.

Table S1: The PRA@PBE, (RPA+rSE)@PBE, and PBE binding energiesfFor the WATER27 dataset. The structural optimization for all the cases uses the specified method with the cc-pVQZ basis sets. Binding energies for RPA@PBE and (RPA+rSE)@PBE are determined at the CBS(5Z,6Z) limit. Differences from the reference values are indicated in parentheses, with negative valuess denoting an overestimation of the binding energy relative to the reference values.[1] Structure relaxation are done in an all-electron manner, while single-point energies are calculated under the frozen-core approximation for the RPA correlation component of the energy and forces.

| System | Ref. [1] | RPA | RPA+rSE | PBE |
|---|---|---|---|---|
| Neutral Molecules | | | | |
| $(H_2O)_2$ | 4.974 | 4.271 (0.703) | 4.878 (0.097) | 6.049 (−1.075) |
| $(H_2O)_3$ | 15.708 | 13.496 (2.212) | 15.700 (0.008) | 19.349 (−3.641) |
| $(H_2O)_4$ | 27.353 | 23.805 (3.548) | 27.432 (−0.079) | 33.827 (−6.474) |
| $(H_2O)_5$ | 35.879 | 31.380 (4.499) | 35.903 (−0.024) | 44.259 (−8.380) |
| $(H_2O)_6$ Prism | 45.988 | 39.793 (6.195) | 46.639 (−0.651) | 55.134 (−9.146) |
| $(H_2O)_6$ Cage | 45.733 | 39.602 (6.131) | 46.230 (−0.497) | 55.255 (−9.522) |
| $(H_2O)_6$ Book | 45.292 | 39.457 (5.835) | 45.550 (−0.258) | 55.444 (−10.152) |
| $(H_2O)_6$ Cyclic | 44.296 | 38.833 (5.463) | 44.136 (0.160) | 54.217 (−9.921) |
| $(H_2O)_8$ d2d | 72.490 | 63.165 (9.325) | 73.477 (−0.987) | 87.003 (−14.513) |
| $(H_2O)_8$ s4 | 72.454 | 63.035 (9.419) | 73.425 (−0.971) | 87.001 (−14.547) |
| $(H_2O)_{20}$ Dodecahedron | 197.804 | 173.365 (24.439) | 200.304 (−2.500) | 239.294 (−41.490) |
| $(H_2O)_{20}$ Edge-sharing | 209.081 | 177.492 (31.589) | 209.501 (−0.420) | 245.352 (−36.271) |
| $(H_2O)_{20}$ Face-cubes | 207.534 | 178.560 (28.975) | 208.201 (−0.667) | 240.356 (−32.822) |
| $(H_2O)_{20}$ Face-sharing | 207.763 | 178.837 (28.926) | 207.740 (0.023) | 242.206 (−34.443) |
| $H_3O^+(H_2O)_6OH^-$ | 268.544 | 259.438 (9.106) | 272.100 (−3.556) | 307.102 (−38.558) |
| Positively Charged Molecules | | | | |
| $H_3O^+H_2O$ | 33.738 | 32.628 (1.110) | 33.962 (−0.224) | 38.560 (−4.822) |
| $H_3O^+(H_2O)_2$ | 57.114 | 54.903 (2.211) | 57.045 (0.069) | 63.559 (−6.445) |
| $H_3O^+(H_2O)_3$ | 76.755 | 73.644 (3.111) | 76.489 (0.266) | 84.117 (−7.362) |
| $H_3O^+(H_2O)_6$ 2d | 114.819 | 108.726 (6.093) | 114.578 (0.241) | 127.542 (−9.859) |
| $H_3O^+(H_2O)_6$ 3d | 117.683 | 110.693 (6.991) | 118.240 (−0.557) | 130.506 (−15.687) |
| Negatively Charged Molecules | | | | |
| $OH^-H_2O$ | 26.687 | 26.123 (0.564) | 27.805 (−1.118) | 39.621 (−12.934) |
| $OH^-(H_2O)_2$ | 48.688 | 47.119 (1.569) | 49.841 (−1.153) | 64.950 (−16.262) |
| $OH^-(H_2O)_3$ | 67.525 | 64.288 (3.237) | 68.754 (−1.229) | 87.148 (−19.623) |
| $OH^-(H_2O)_4$ c4 | 84.351 | 80.172 (4.179) | 86.030 (−1.679) | 105.844 (−21.493) |
| $OH^-(H_2O)_4$ cs | 85.015 | 80.029 (4.986) | 86.456 (−1.441) | 107.872 (−22.857) |
| $OH^-(H_2O)_5$ | 100.782 | 94.131 (6.651) | 102.429 (−1.647) | 124.846 (−24.064) |
| $OH^-(H_2O)_6$ | 115.672 | 107.662 (8.010) | 117.420 (−1.748) | 142.308 (−26.636) |

Table S2: The same as Table S1, with structures optimized using the cc-pV5Z basis sets

| System | Ref. [1] | RPA | RPA+rSE | PBE |
|---|---|---|---|---|
| Neutral Molecules | | | | |
| $(H_2O)_2$ | 4.974 | 4.284 (0.690) | 4.900 (0.074) | 5.375 (−0.401) |
| $(H_2O)_3$ | 15.708 | 13.552 (2.156) | 15.853 (−0.145) | 17.064 (−1.356) |
| $(H_2O)_4$ | 27.353 | 23.935 (3.419) | 27.581 (−0.228) | 30.512 (−3.159) |
| $(H_2O)_5$ | 35.879 | 31.497 (4.382) | 36.085 (−0.206) | 40.248 (−4.369) |
| $(H_2O)_6$ Prism | 45.988 | 39.810 (6.176) | 46.629 (−0.641) | 48.868 (−2.880) |
| $(H_2O)_6$ Cage | 45.733 | 39.657 (6.076) | 46.388 (−0.655) | 49.203 (−3.470) |
| $(H_2O)_6$ Book | 45.292 | 39.550 (5.743) | 45.701 (−0.409) | 49.967 (−4.675) |
| $(H_2O)_6$ Cyclic | 44.296 | 38.945 (5.351) | 44.378 (−0.082) | 49.541 (−5.245) |
| $(H_2O)_8$ d2d | 72.490 | 63.049 (9.441) | 73.401 (−0.911) | 78.042 (−5.552) |
| $(H_2O)_8$ s4 | 72.454 | 63.088 (9.366) | 73.401 (−0.947) | 78.049 (−5.595) |
| $(H_2O)_{20}$ Dodecahedron | 197.804 | 173.269 (24.535) | 200.208 (−2.404) | 217.602 (−19.798) |
| $(H_2O)_{20}$ Edge-sharing | 209.081 | 177.396 (31.685) | 209.406 (−0.325) | 221.307 (−12.226) |
| $(H_2O)_{20}$ Face-cubes | 207.534 | 178.464 (29.070) | 208.105 (−0.571) | 215.682 (−8.148) |
| $(H_2O)_{20}$ Face-sharing | 207.763 | 178.742 (29.021) | 207.644 (0.119) | 217.968 (−10.205) |
| $H_3O^+(H_2O)_8OH^-$ | 268.544 | 259.191 (9.353) | 271.393 (−2.849) | 290.085 (−21.541) |
| Positively Charged Molecules | | | | |
| $H_3O^+H_2O$ | 33.738 | 32.637 (1.101) | 33.939 (−0.201) | 37.391 (−3.653) |
| $H_3O^+(H_2O)_2$ | 57.114 | 54.945 (2.169) | 57.082 (0.0322) | 61.457 (−4.343) |
| $H_3O^+(H_2O)_3$ | 76.755 | 73.675 (3.080) | 76.528 (0.227) | 81.135 (−4.380) |
| $H_3O^+(H_2O)_6$ 2d | 114.819 | 108.865 (5.954) | 114.712 (0.107) | 121.785 (−6.966) |
| $H_3O^+(H_2O)_6$ 3d | 117.683 | 110.741 (6.943) | 118.244 (−0.561) | 123.777 (−6.094) |
| Negatively Charged Molecules | | | | |
| $OH^-H_2O$ | 26.687 | 26.194 (0.493) | 27.958 (−1.271) | 34.520 (−7.833) |
| $OH^-(H_2O)_2$ | 48.688 | 47.214 (1.474) | 50.028 (−1.340) | 57.968 (−9.280) |
| $OH^-(H_2O)_3$ | 67.525 | 64.510 (3.015) | 69.007 (−1.482) | 77.806 (−10.281) |
| $OH^-(H_2O)_4$ c4 | 84.351 | 80.241 (4.110) | 86.207 (−1.856) | 93.560 (−9.209) |
| $OH^-(H_2O)_4$ cs | 85.015 | 80.124 (4.891) | 86.621 (−1.606) | 95.457 (−10.442) |
| $OH^-(H_2O)_5$ | 100.782 | 94.210 (6.572) | 102.531 (−1.749) | 110.572 (−9.790) |
| $OH^-(H_2O)_6$ | 115.672 | 107.644 (8.028) | 117.537 (−1.865) | 126.615 (−10.943) |

Table S3: The same as Table S1, with structures optimized using the NAO-VCC-4Z basis sets, except for $(H_2O)_{20}$ whose structures are optimized using the NAO-VCC-3Z basis sets.

| System | Ref. [1] | RPA | RPA+rSE | PBE |
|---|---|---|---|---|
| Neutral Molecules | | | | |
| $(H_2O)_2$ | 4.974 | 4.223 (0.751) | 4.849 (0.125) | 5.141 (−0.167) |
| $(H_2O)_3$ | 15.708 | 13.429 (2.279) | 15.713 (−0.005) | 16.469 (−0.761) |
| $(H_2O)_4$ | 27.353 | 23.633 (3.720) | 27.394 (−0.041) | 29.624 (−2.271) |
| $(H_2O)_5$ | 35.879 | 31.255 (4.624) | 35.836 (0.043) | 39.112 (−3.233) |
| $(H_2O)_6$ Prism | 45.988 | 39.429 (6.559) | 46.294 (−0.306) | 47.314 (−1.326) |
| $(H_2O)_6$ Cage | 45.733 | 39.327 (6.406) | 46.077 (−0.344) | 47.700 (−1.967) |
| $(H_2O)_6$ Book | 45.292 | 39.303 (5.989) | 45.435 (−0.143) | 48.529 (−3.237) |
| $(H_2O)_6$ Cyclic | 44.296 | 38.519 (5.777) | 44.086 (0.210) | 48.148 (−3.852) |
| $(H_2O)_8$ d2d | 72.490 | 62.475 (10.015) | 72.956 (−0.466) | 75.842 (−3.352) |
| $(H_2O)_8$ s4 | 72.454 | 62.743 (9.711) | 73.094 (−0.640) | 75.840 (−3.386) |
| $(H_2O)_{20}$ Dodecahedron | 197.804 | 172.072 (25.732) | 199.241 (−1.4374) | 211.902 (−14.098) |
| $(H_2O)_{20}$ Edge-sharing | 209.081 | 179.670 (29.411) | 209.158 (−0.077) | 215.446 (−6.365) |
| $(H_2O)_{20}$ Face-cubes | 207.534 | 176.565 (30.969) | 207.625 (−0.091) | 209.763 (−2.229) |
| $(H_2O)_{20}$ Face-sharing | 207.763 | 177.101 (30.662) | 207.318 (0.445) | 212.071 (−4.308) |
| $H_3O^+(H_2O)_8OH^-$ | 268.544 | 258.978 (9.566) | 271.732 (−3.188) | 288.251 (−19.707) |
| Positively Charged Molecules | | | | |
| $H_3O^+H_2O$ | 33.738 | 32.614 (1.123) | 33.931 (−0.193) | 37.141 (−3.403) |
| $H_3O^+(H_2O)_2$ | 57.114 | 54.841 (2.273) | 56.995 (0.119) | 60.975 (−3.861) |
| $H_3O^+(H_2O)_3$ | 76.755 | 73.543 (3.212) | 76.443 (0.312) | 80.432 (−3.677) |
| $H_3O^+(H_2O)_6$ 2d | 114.819 | 108.597 (6.222) | 114.393 (0.426) | 120.362 (−5.543) |
| $H_3O^+(H_2O)_6$ 3d | 117.683 | 110.486 (7.197) | 118.030 (−0.347) | 122.184 (−4.501) |
| Negatively Charged Molecules | | | | |
| $OH^-H_2O$ | 26.687 | 26.210 (0.477) | 27.919 (−1.232) | 34.524 (−7.837) |
| $OH^-(H_2O)_2$ | 48.688 | 47.142 (1.546) | 49.613 (−0.925) | 57.891 (−9.203) |
| $OH^-(H_2O)_3$ | 67.525 | 64.266 (3.258) | 68.785 (−1.260) | 77.186 (−9.661) |
| $OH^-(H_2O)_4$ c4 | 84.351 | 80.076 (4.275) | 86.051 (−1.700) | 92.553 (−8.202) |
| $OH^-(H_2O)_4$ cs | 85.015 | 80.017 (4.998) | 86.516 (−1.501) | 94.328 (−9.313) |
| $OH^-(H_2O)_5$ | 100.782 | 94.129 (6.653) | 102.402 (−1.620) | 109.106 (−8.324) |
| $OH^-(H_2O)_6$ | 115.672 | 107.565 (8.106) | 117.449 (−1.777) | 124.830 (−9.158) |

## 2  RPA+rSE Structures for the WATER27 Dataset

The atomic coordinates for the WATER27 dataset provided below are obtained using the (RPA+rSE)@PBE method with the cc-pVQZ basis sets. Namely, these geometries are used in calculating the binding energies reported in Table S1.

- **$H_2O$**

$$
\begin{array}{cccc}
-0.12542435 & 0.00000008 & 0.00000000 & O \\
0.47241421 & 0.75543439 & 0.00000000 & H \\
0.47241427 & -0.75543447 & 0.00000000 & H \\
\end{array}
$$

- **$(H_2O)_2$**

$$
\begin{array}{cccc}
0.19117878 & -1.21038300 & 0.77500413 & O \\
-0.14942899 & -1.02434545 & 1.71840345 & H \\
-0.05241612 & -0.34300708 & 0.36445185 & H \\
0.19473091 & 1.22723126 & -0.72984596 & O \\
-0.58005884 & 1.23348829 & -1.30410233 & H \\
0.91636397 & 0.98527219 & -1.32214019 & H \\
\end{array}
$$

- **$(H_2O)_3$**

|  |  |  |  |
|---:|---:|---:|:--|
| 1.17185028 | 1.08977476 | −0.02250623 | O |
| 0.45522957 | 0.91780742 | 0.62022273 | H |
| 1.98429900 | 0.95106606 | 0.47525062 | H |
| −1.13410745 | 0.06296289 | 1.12838419 | O |
| −1.13482594 | −0.54671256 | 1.87334586 | H |
| −1.00769656 | −0.50855139 | 0.34432217 | H |
| −0.10771248 | −1.07251867 | −1.17753691 | O |
| 0.52362046 | −0.34715044 | −0.99741541 | H |
| −0.52532375 | −0.82485535 | −2.00905649 | H |

- $(H_2O)_4$

|  |  |  |  |
|---:|---:|---:|:--|
| −1.42004769 | 1.24115954 | −0.34325813 | O |
| −0.45994337 | 1.42830720 | −0.22410650 | H |
| −1.85089163 | 1.73691563 | 0.36077713 | H |
| −1.26504212 | −1.44019985 | 0.03189874 | O |
| −1.43601974 | −0.47903179 | −0.10204494 | H |
| −1.58968764 | −1.85665042 | −0.77323275 | H |
| 1.43128132 | −1.25014085 | 0.25164456 | O |
| 0.46657847 | −1.43367603 | 0.16986711 | H |
| 1.66842020 | −1.59213136 | 1.11990310 | H |
| 1.25363436 | 1.44911306 | 0.05964029 | O |
| 1.42917804 | 0.48433663 | 0.15665040 | H |
| 1.77178808 | 1.71177059 | −0.70820480 | H |

- $(H_2O)_5$

|  |  |  |  |
|---:|---:|---:|:---:|
| 1.16783153 | −1.50763868 | 1.25425110 | O |
| −0.76759075 | −1.98413672 | 1.98900993 | H |
| −1.04833503 | −0.55625505 | 1.48777013 | H |
| 0.10000095 | −2.01971656 | −1.06541376 | O |
| −0.55547601 | −2.46778350 | −1.60963655 | H |
| −0.36702549 | −1.84517121 | −0.21376991 | H |
| 0.79853089 | 2.14161344 | −0.04436100 | O |
| 0.93605482 | 1.47490680 | −0.75861042 | H |
| 1.68626617 | 2.34410299 | 0.26838132 | H |
| −0.77248785 | 1.08231445 | 1.86492698 | O |
| −0.19027243 | 1.48681864 | 1.17836349 | H |
| −1.57237077 | 1.61764058 | 1.84799752 | H |
| 1.15938167 | 0.29534018 | −1.97124930 | O |
| 0.69245268 | 0.52985913 | −2.77975142 | H |
| 0.76057500 | −0.56159179 | −1.69032320 | H |

- **Prism: $(H_2O)_6$**

|             |             |             |   |
|------------:|------------:|------------:|---|
|  1.45938166 | −0.35424892 |  1.37745511 | O |
|  0.52072990 | −0.68298832 |  1.49146648 | H |
|  1.95375216 | −0.68251848 |  2.13436096 | H |
|  0.76386212 |  1.95265891 | −0.06879508 | O |
|  1.09017973 |  1.40420849 |  0.66633221 | H |
|  1.12186714 |  1.47477089 | −0.83322034 | H |
|  1.62862383 | −0.46965541 | −1.33533754 | O |
|  2.35240596 | −0.91831897 | −1.78343710 | H |
|  1.77145335 | −0.61870376 | −0.37704474 | H |
| −1.83493941 |  1.20170110 | −0.05541295 | O |
| −0.92880402 |  1.59223354 | −0.07491749 | H |
| −2.43817743 |  1.94429362 | −0.15653002 | H |
| −1.11800612 | −1.25941579 | −1.34461633 | O |
| −0.18152061 | −1.05960515 | −1.51676753 | H |
| −1.51441704 | −0.38054138 | −1.22021144 | H |
| −1.05519551 | −1.10387805 |  1.40545810 | O |
| −1.10038256 | −1.42755165 |  0.47785061 | H |
| −1.52757493 | −0.25956551 |  1.31584072 | H |

- **Cage: $(\mathbf{H_2O})_6$**

9

|            |            |            |   |
|-----------:|-----------:|-----------:|---|
| $-2.05513340$ | $0.95569733$ | $1.71949088$ | O |
| $-1.93788018$ | $1.46883060$ | $2.52543194$ | H |
| $-1.50914692$ | $1.41718850$ | $1.04380909$ | H |
| $-0.59888588$ | $-1.30754841$ | $1.15695282$ | O |
| $-1.15983707$ | $-0.58954646$ | $1.52118379$ | H |
| $0.30102327$ | $-0.94364732$ | $1.22475946$ | H |
| $1.78768770$ | $0.14780743$ | $0.66397871$ | O |
| $1.99468509$ | $-0.19931830$ | $-0.23671274$ | H |
| $2.63429660$ | $0.24559241$ | $1.11025665$ | H |
| $-0.80009979$ | $-0.83362920$ | $-1.42271214$ | O |
| $-1.49146322$ | $-1.33458452$ | $-1.86573864$ | H |
| $-0.81659366$ | $-1.12091353$ | $-0.46413731$ | H |
| $-0.37379627$ | $1.76177282$ | $-0.29091068$ | O |
| $-0.65346966$ | $1.04061585$ | $-0.88031972$ | H |
| $0.45506401$ | $1.41717154$ | $0.08446550$ | H |
| $1.87829714$ | $-0.77846165$ | $-1.86026888$ | O |
| $2.03906961$ | $-0.07493845$ | $-2.49882432$ | H |
| $0.90267961$ | $-0.89505842$ | $-1.86894298$ | H |

- **Book: $(\mathbf{H_2O})_6$**

|  |  |  |  |
|---:|---:|---:|:---:|
| 0.11089988 | 1.51705936 | 0.81408095 | O |
| 0.96475686 | 1.47700110 | 0.30767394 | H |
| 0.27196299 | 2.11952494 | 1.54697093 | H |
| −0.02945623 | −1.33718430 | 0.95869321 | O |
| −0.03234359 | −0.37780624 | 1.11514554 | H |
| −0.86075894 | −1.48495865 | 0.46151014 | H |
| −2.33438416 | 1.35286220 | −0.42657987 | O |
| −1.45722478 | 1.51657417 | −0.02463810 | H |
| −2.29025210 | 1.77267237 | −1.29185694 | H |
| −2.40547940 | −1.39399576 | −0.45551229 | O |
| −2.47148447 | −0.41633297 | −0.50127882 | H |
| −3.15547799 | −1.65726333 | 0.08851707 | H |
| 2.40800083 | 1.18083418 | −0.45934557 | O |
| 2.44298462 | 0.18968158 | −0.47973345 | H |
| 2.44947031 | 1.44359612 | −1.38476069 | H |
| 2.22142711 | −1.47452353 | −0.45598992 | O |
| 1.36316074 | −1.53167911 | 0.03909078 | H |
| 2.84699893 | −1.96620510 | 0.08607398 | H |

- **Cyclic: $(H_2O)_6$**

|           |            |            |   |
|-----------|------------|------------|---|
| 1.84373752 | 1.94528868 | 0.06905276 | O |
| 0.87921567 | 2.13988113 | 0.00028964 | H |
| 2.11280758 | 2.36937574 | 0.89012773 | H |
| 0.76313790 | −2.56977056 | 0.06900593 | O |
| 0.99560247 | −3.01459485 | 0.89028920 | H |
| 1.41399335 | −1.83182153 | 0.00026638 | H |
| 2.60737110 | −0.62431897 | −0.06906277 | O |
| 2.29327264 | 0.30814936 | −0.00022661 | H |
| 3.10861252 | −0.64514253 | −0.89048934 | H |
| −1.84392642 | −1.94544538 | −0.06898525 | O |
| −2.11278393 | −2.36937535 | −0.89021243 | H |
| −0.87937824 | −2.13987613 | −0.00018256 | H |
| −0.76314621 | 2.56989597 | −0.06896784 | O |
| −0.99550556 | 3.01459959 | −0.89034684 | H |
| −1.41393442 | 1.83188366 | −0.00026116 | H |
| −2.60720920 | 0.62429657 | 0.06904453 | O |
| −3.10857149 | 0.64517842 | 0.89039576 | H |
| −2.29321247 | −0.30820954 | 0.00025032 | H |

- **d2d: $(H_2O)_8$**

|  |  |  |  |
|---:|---:|---:|:---:|
| 1.41700989 | −1.41654977 | 1.33669086 | O |
| −1.41662692 | −1.41689634 | −1.33652212 | O |
| 1.41678491 | 1.41673333 | −1.33647814 | O |
| −1.41702163 | 1.41667275 | 1.33632671 | O |
| 1.48395576 | −1.48366740 | 0.34285478 | H |
| −1.48361027 | −1.48386700 | −0.34267412 | H |
| 1.48373638 | 1.48382369 | −0.34263952 | H |
| −1.48386359 | 1.48362734 | 0.34246603 | H |
| 2.04829704 | −2.04823107 | 1.69405511 | H |
| −2.04794850 | −2.04859223 | −1.69380319 | H |
| 2.04813624 | 2.04836753 | −1.69381712 | H |
| −2.04854191 | 2.04823501 | 1.69349445 | H |
| 1.37106154 | −1.37098829 | −1.29983708 | O |
| −1.37082385 | −1.37101235 | 1.29994120 | O |
| 1.37076065 | 1.37114639 | 1.30004657 | O |
| −1.37092441 | 1.37081735 | −1.30012140 | O |
| 0.42148410 | −1.50620877 | −1.48000603 | H |
| −0.42122511 | −1.50598547 | 1.48017800 | H |
| 0.42112983 | 1.50614789 | 1.48009953 | H |
| 1.50624862 | −0.42141396 | −1.48003149 | H |
| −1.50610515 | 0.42123663 | −1.48030069 | H |
| 1.50615408 | 0.42160990 | 1.48028120 | H |
| −1.50625901 | −0.42146174 | 1.48009221 | H |
| −0.42133275 | 1.50602727 | −1.48022674 | H |

- **s4: $(H_2O)_8$**

|          |            |            |   |
|---------:|-----------:|-----------:|---|
|  1.94711924 | −0.08399369 | −1.42229182 | O |
|  1.28022100 | −0.82502134 | −1.48019502 | H |
|  2.65217456 | −0.30720314 | −2.03763277 | H |
| −0.08258401 | −1.94845802 |  1.42234024 | O |
| −0.30502697 | −2.65557479 |  2.03556417 | H |
| −0.82433202 | −1.28262812 |  1.48092967 | H |
| −1.88553917 | −0.03014873 |  1.36518212 | O |
| −2.10512838 |  0.02345636 |  0.41557710 | H |
| −1.33487237 |  0.76119901 |  1.51600286 | H |
|  0.02849877 | −1.88634233 | −1.36489213 | O |
| −0.02448032 | −2.10562191 | −0.41519851 | H |
| −0.76247874 | −1.33502881 | −1.51532974 | H |
| −0.02854474 |  1.88634737 | −1.36486027 | O |
|  0.76244248 |  1.33505410 | −1.51531345 | H |
|  0.02441703 |  2.10560129 | −0.41515681 | H |
|  0.08259924 |  1.94848096 |  1.42235678 | O |
|  0.82434228 |  1.28263550 |  1.48089186 | H |
|  0.30508259 |  2.65556718 |  2.03560255 | H |
| −1.94707772 |  0.08395005 | −1.42236459 | O |
| −1.28022776 |  0.82502231 | −1.48022779 | H |
| −2.65216668 |  0.30714324 | −2.03767429 | H |
|  1.88552814 |  0.03016931 |  1.36517219 | O |
|  2.10520254 | −0.02343550 |  0.41558630 | H |
|  1.33483101 | −0.76117031 |  1.51593133 | H |

- **Dodecahedron: $(H_2O)_{20}$**

| | | | |
|---:|---:|---:|:--|
| 1.56256649 | 3.51110382 | 0.09121125 | O |
| 3.38930534 | 1.50785703 | 0.85485725 | O |
| −0.55169748 | 3.37498506 | 1.73188637 | O |
| 0.62246303 | 2.85959819 | −2.45360532 | O |
| 2.46886286 | 0.22375134 | 2.89334040 | O |
| −0.01263598 | 1.28423185 | 3.55386362 | O |
| 3.60850421 | −0.20042875 | −1.20613447 | O |
| −2.62360390 | 2.69258707 | 0.34714796 | O |
| 1.89127278 | 0.57576377 | −3.25822841 | O |
| −1.92589823 | 2.40825700 | −2.28022199 | O |
| 1.99568940 | −2.36452135 | 2.16353599 | O |
| −1.84756892 | −0.63828330 | 3.22105420 | O |
| 2.70081067 | −2.70928761 | −0.28125575 | O |
| −3.59660306 | 0.22766637 | 1.18743963 | O |
| 0.11451014 | −1.27677410 | −3.57020698 | O |
| −2.31881908 | −0.26083282 | −3.00574650 | O |
| −0.70492622 | −2.81346071 | 2.46711658 | O |
| 0.55436385 | −3.35641997 | −1.71940730 | O |
| −3.36639998 | −1.59769302 | −0.74882642 | O |
| −1.65073445 | −3.38398969 | −0.03243388 | O |
| 0.79488083 | 3.53644794 | 0.71246431 | H |
| 1.95589982 | 4.38945178 | 0.12892862 | H |
| 2.76449872 | 2.19912704 | 0.57541560 | H |
| 3.47407242 | 0.90301253 | 0.07757152 | H |
| −0.83563504 | 4.14876137 | 2.23043970 | H |
| −1.35883978 | 3.10016365 | 1.19494777 | H |
| 0.98839162 | 3.07614758 | −1.57626527 | H |
| 1.10125764 | 2.05539562 | −2.74576586 | H |
| 3.13077044 | 0.33529195 | 3.58388685 | H |

| | | | |
|---:|---:|---:|---|
| 2.83266497 | 0.72577921 | 2.09701816 | H |
| 0.87438660 | 0.94185780 | 3.32980406 | H |
| −0.16762033 | 2.00801562 | 2.92186759 | H |
| 3.28340537 | −1.07006700 | −0.91337790 | H |
| 3.01947588 | 0.04959136 | −1.94343990 | H |
| −2.40560880 | 2.60899079 | −0.60489917 | H |
| −2.97659973 | 1.82098369 | 0.60765933 | H |
| 1.19876644 | −0.14551407 | −3.38227926 | H |
| 2.33171113 | 0.64827987 | −4.11164386 | H |
| −2.33405649 | 3.02001374 | −2.90203314 | H |
| −0.93831984 | 2.57932059 | −2.35469300 | H |
| 1.03825936 | −2.50805295 | 2.30550312 | H |
| 2.16753883 | −1.44317701 | 2.45212645 | H |
| −2.45272367 | −0.31071304 | 2.53317347 | H |
| −1.18062386 | 0.08190131 | 3.34672118 | H |
| 3.45518482 | −3.30639304 | −0.23442444 | H |
| 2.41829934 | −2.56794923 | 0.68006073 | H |
| −4.50861275 | 0.26270668 | 1.49523223 | H |
| −3.58254578 | −0.46335549 | 0.47994271 | H |
| 0.25420800 | −2.00062250 | −2.93529615 | H |
| −0.78645303 | −0.92635020 | −3.36942661 | H |
| −2.20785849 | 0.67430126 | −2.75102862 | H |
| −2.68393996 | −0.69344528 | −2.21411454 | H |
| −0.89868362 | −3.45356862 | 3.16014355 | H |
| −1.15501880 | −1.95806427 | 2.76591787 | H |
| 0.69489525 | −4.20714631 | −2.14868663 | H |
| 1.38214564 | −3.17951620 | −1.20562484 | H |
| −4.13477066 | −2.10154987 | −1.03811098 | H |
| −2.68840793 | −2.29158316 | −0.47144344 | H |
| 0.85978700 | 3.38390870 | 0.60513323 | H |

- **Edge-Sharing: $(H_2O)_{20}$**

$$
\begin{array}{rrrl}
1.40542915 & -2.21275368 & 1.30252298 & O \\
0.89768010 & -1.37629120 & 1.42056609 & H \\
2.32702552 & -1.97484649 & 1.52462974 & H \\
1.36563458 & -2.40405846 & -1.37785962 & O \\
1.39074123 & -2.41665710 & -0.38922770 & H \\
0.69093776 & -3.07127998 & -1.60872327 & H \\
-0.00182666 & 0.01653064 & 1.40046022 & O \\
0.36580094 & 0.92735319 & 1.49776701 & H \\
-0.98364398 & 0.04890717 & 1.49560623 & H \\
-2.64355864 & -0.08665325 & 1.31504927 & O \\
-2.67946454 & -0.09595649 & 0.32649081 & H \\
-2.85084221 & -1.00712577 & 1.57083475 & H \\
3.90164143 & 1.43818230 & -1.41841759 & O \\
4.64060242 & 1.71109315 & -1.97115508 & H \\
3.86484494 & 0.44205518 & -1.48399205 & H \\
-0.96316192 & 4.09234592 & 1.46150601 & O \\
-1.14016581 & 4.85354939 & 2.02281904 & H \\
-1.78744539 & 3.52771028 & 1.49192126 & H \\
-3.10905873 & 2.64337673 & -1.50892431 & O \\
-3.69536892 & 3.10412283 & -2.11731437 & H \\
\end{array}
$$

| | | | |
|---:|---:|---:|:--|
| −2.24016062 | 3.13783094 | −1.53739654 | H |
| 0.07539021 | −0.08573294 | −1.26060119 | O |
| 0.54908133 | −0.92863501 | −1.46321786 | H |
| 0.06160013 | −0.07197258 | −0.26971217 | H |
| 1.30903528 | 2.25780785 | −1.36175527 | O |
| 2.25574639 | 2.09867309 | −1.54374126 | H |
| 0.89156035 | 1.36752995 | −1.45340158 | H |
| 3.93401027 | −1.23461651 | 1.45754623 | O |
| 4.65738226 | −1.46886096 | 2.04734692 | H |
| 3.83741003 | −0.24210066 | 1.52188349 | H |
| −0.84740756 | 3.95619942 | −1.34784268 | O |
| −0.81922709 | 4.17719085 | −0.39852227 | H |
| −0.03646904 | 3.43305618 | −1.49783243 | H |
| −2.57570952 | −0.02352372 | −1.37035353 | O |
| −2.83625391 | 0.89284775 | −1.58893590 | H |
| −1.59370256 | −0.04492076 | −1.46550387 | H |
| −3.01840681 | −2.73131503 | −1.28132050 | O |
| −2.97315020 | −1.77400677 | −1.46586414 | H |
| −3.18092547 | −2.77940745 | −0.32070324 | H |
| 1.14817040 | 2.39559963 | 1.32133523 | O |
| 0.48511237 | 3.07880925 | 1.54253618 | H |
| 1.20560593 | 2.41629225 | 0.33446093 | H |
| −0.77914351 | −4.08279795 | −1.50401474 | O |
| −0.94659677 | −4.85658045 | −2.05112778 | H |
| −1.62686342 | −3.55364410 | −1.51234311 | H |
| −3.12688960 | 2.62524479 | 1.29769193 | O |
| −3.04812870 | 1.67405820 | 1.49918088 | H |
| −3.30550641 | 2.64831583 | 0.33860186 | H |
| 3.85218437 | −1.18455213 | −1.34781053 | O |
| 3.03142537 | 1.66522989 | 1.56010749 | H |

- **Face-Cubes: $(H_2O)_{20}$**

|  |  |  |  |
|---:|---:|---:|:---|
| 0.00702370 | 1.89532569 | −5.39821423 | $O$ |
| −1.95584631 | −0.10711447 | −5.51364078 | $O$ |
| −0.12184593 | 1.94732268 | 5.51694240 | $O$ |
| −1.87755272 | −0.01677627 | 5.40620090 | $O$ |
| −1.89454014 | −0.00984364 | 2.63150484 | $O$ |
| 1.88743344 | −0.00116354 | 2.62814523 | $O$ |
| −1.90641704 | 0.01024101 | −2.79031460 | $O$ |
| 1.90443302 | 0.00100493 | −2.78774400 | $O$ |
| 0.00252655 | −1.86939532 | −5.40845577 | $O$ |
| 1.96577419 | 0.13402397 | −5.50882337 | $O$ |
| 0.13172140 | −1.97365133 | 5.50541621 | $O$ |
| 1.88700272 | −0.00956453 | 5.40050669 | $O$ |

| | | | |
|---:|---:|---:|:---|
| 1.89666616 | 0.09216035 | −0.08543788 | O |
| −0.00179868 | 1.89618397 | −2.62469698 | O |
| −0.00234174 | 1.89938392 | 2.79402485 | O |
| −0.00421876 | −1.91135666 | 2.78419580 | O |
| −0.09708224 | 1.90193249 | 0.09042728 | O |
| 0.08627161 | −1.90215340 | 0.08069720 | O |
| −0.00021549 | −1.88557338 | −2.63471988 | O |
| −1.90771207 | −0.09117362 | −0.08578102 | O |
| −0.77884557 | 1.33349324 | −5.53466709 | H |
| −0.04957756 | 2.16334344 | −4.46350711 | H |
| −1.27749343 | −0.83856998 | −5.54756822 | H |
| −2.60828834 | −0.30679087 | −6.19204361 | H |
| −0.32487861 | 2.59853016 | 6.19553913 | H |
| −0.85116515 | 1.26662738 | 5.54883731 | H |
| −1.31297343 | −0.80167408 | 5.53664400 | H |
| −2.15063299 | −0.06993594 | 4.47279731 | H |
| −2.24243189 | −0.03060235 | 1.72276391 | H |
| −1.26768597 | 0.75321716 | 2.64534843 | H |
| 2.23279667 | 0.02418048 | 1.71844666 | H |
| 1.26024104 | −0.76400237 | 2.63938220 | H |
| −2.19575865 | −0.02993945 | −3.72104087 | H |
| −1.27723991 | 0.77141090 | −2.75526580 | H |
| 1.27524738 | −0.76041137 | −2.75804511 | H |
| 2.19689330 | 0.04602234 | −3.71730380 | H |
| 0.05611732 | −2.14234612 | −4.47500028 | H |
| 0.78903996 | −1.30733238 | −5.53991663 | H |
| 1.28768529 | 0.86587264 | −5.54030433 | H |
| 2.62159112 | 0.33858065 | −6.18249078 | H |
| 0.86133018 | −1.29316102 | 5.53882419 | H |
| 0.33774996 | 2.62978691 | 6.17831261 | H |

- **Face-Sharing: $(H_2O)_{20}$**

$$
\begin{array}{rrrl}
-0.44407378 & -1.32182539 & 2.25302051 & O \\
-0.46483370 & -0.41549326 & 2.60244188 & H \\
-1.12551097 & -1.31667272 & 1.53586404 & H \\
-1.16009683 & -4.23523279 & -1.97737168 & O \\
-1.61019517 & -4.89305746 & -2.51641282 & H \\
\end{array}
$$

| | | | |
|---:|---:|---:|:---:|
| −1.61569787 | −4.24811991 | −1.07903316 | H |
| −0.39215866 | −4.13396201 | 2.29659853 | O |
| −0.39324974 | −3.19992657 | 2.56419694 | H |
| 0.48783418 | −4.26560454 | 1.89997551 | H |
| −2.28546090 | −1.29929499 | 0.31229366 | O |
| −2.62625458 | −0.39034376 | 0.35343644 | H |
| −1.80014663 | −1.34826805 | −0.54828123 | H |
| −1.00151110 | −1.52936561 | −2.02134512 | O |
| −0.03061856 | −1.48780725 | −1.83552407 | H |
| −1.16696082 | −2.46839910 | −2.23778042 | H |
| 1.58500978 | −4.14519164 | −1.57520527 | O |
| 0.63113408 | −4.27581673 | −1.74642867 | H |
| 1.73936039 | −3.20924017 | −1.79297331 | H |
| −2.29108247 | −4.09366379 | 0.35133047 | O |
| −1.61575598 | −4.18529695 | 1.06497480 | H |
| −2.57469774 | −3.16713282 | 0.42625947 | H |
| 1.61259258 | −1.36722620 | −1.53925086 | O |
| 1.72433499 | −1.39523494 | −0.55577752 | H |
| 1.85717814 | −0.46094402 | −1.79307912 | H |
| 1.94058283 | −1.54518578 | 1.09752657 | O |
| 1.04761923 | −1.49219606 | 1.52243411 | H |
| 2.20835314 | −2.48063474 | 1.19244338 | H |
| 2.11063967 | −4.24686005 | 0.99588734 | O |
| 2.80866469 | −4.88603359 | 1.16901069 | H |
| 1.93042347 | −4.28971469 | 0.01404387 | H |
| −0.36459588 | 4.20057917 | 2.24363904 | O |
| −0.51074386 | 4.87767608 | 2.91210260 | H |
| −1.07168239 | 4.32546912 | 1.56842847 | H |
| −1.07858269 | 1.16913065 | −1.98954380 | O |
| −1.21998790 | 0.25065813 | −2.28440196 | H |

- $H_3O^+(H_2O)_6OH^-$

|  |  |  |  |
|---:|---:|---:|:---:|
| −1.07456569 | 1.38052369 | −1.53951656 | O |
| −1.12100777 | 1.38161293 | 0.97575458 | O |
| 1.36293029 | 1.30331737 | 1.36837045 | O |
| −1.60285240 | −1.06426533 | 1.31472005 | O |
| 0.85040941 | −1.60002215 | 1.43119943 | O |
| 1.21688929 | −1.50026416 | −1.05915539 | O |
| 1.80799026 | 0.94814079 | −1.08007254 | O |
| −1.21318588 | −1.12861561 | −1.60874994 | O |
| −1.18021129 | 0.35414717 | −1.64948956 | H |
| −1.76687145 | 1.79483810 | −2.06432676 | H |
| −1.63242229 | −1.38604889 | −0.77500435 | H |
| −2.08016965 | −1.39003684 | 2.08440478 | H |
| −0.61282173 | −1.35558987 | 1.42299628 | H |
| −0.22244097 | −1.37811315 | −1.46563023 | H |
| −1.16978097 | 1.47047656 | −0.05237105 | H |
| −1.40266287 | 0.40939727 | 1.18240590 | H |
| 1.03999442 | 1.31960045 | −1.53762677 | H |
| 1.08047471 | −1.65739001 | 0.42727706 | H |
| −0.11629133 | 1.43369314 | 1.20988971 | H |
| 1.27903369 | −0.77911727 | 1.71340077 | H |
| 1.74750102 | −2.15441265 | −1.52116744 | H |
| 1.66146111 | −0.07045812 | −1.15411259 | H |
| 1.85036171 | 2.05567358 | 1.71869400 | H |
| 1.62772052 | 1.20522032 | 0.36983217 | H |

- $H_3O^+$

|           |            |            |   |
|-----------|------------|------------|---|
| 0.06646323 | −0.08246581 | −0.03348625 | O |
| −0.55691291 | 0.46622945 | 0.61498715 | H |
| 0.18275114 | 0.45176050 | −0.81746575 | H |
| 0.73009773 | −0.47635410 | 0.38180988 | H |

- **$H_3O^+(H_2O)$**

|           |            |            |   |
|-----------|------------|------------|---|
| −1.19933530 | 0.05430154 | 0.07438896 | O |
| −1.61719686 | −0.73375241 | 0.45564533 | H |
| −1.67477957 | 0.28091715 | −0.73962141 | H |
| 1.19937340 | −0.05439091 | 0.07433622 | O |
| 0.00012115 | −0.00015223 | 0.00177180 | H |
| 1.67479934 | −0.28061245 | −0.73979819 | H |
| 1.61699773 | 0.73368364 | 0.45581933 | H |

- **$H_3O^+(H_2O)_2$**

|           |            |            |   |
|-----------|------------|------------|---|
| 1.24449979 | 1.38306436 | 0.99602772 | O |
| 1.78365802 | 1.17168899 | 1.76839991 | H |
| 1.22545150 | 2.34692187 | 0.93573836 | H |
| −0.28549962 | −1.45637849 | −1.49985290 | O |
| −0.13920376 | −2.40215460 | −1.37322491 | H |
| −0.77680245 | −1.36753832 | −2.32673749 | H |
| −0.81255719 | 0.15882790 | 0.31770720 | O |
| 0.04156684 | 0.64554960 | 0.66335531 | H |
| −0.61168603 | −0.56707836 | −0.40243162 | H |
| −1.33016486 | −0.20587691 | 1.04990946 | H |

- **$H_3O^+(H_2O)_3$**

|  |  |  |  |
|---|---|---|---|
| 0.20098470 | 1.28725718 | −2.07960805 | O |
| 0.72323055 | 2.07772501 | −2.26490326 | H |
| 0.25280609 | 0.75116001 | −2.88045704 | H |
| −1.91681036 | 0.38933499 | 1.48330756 | O |
| −2.76422120 | 0.47625164 | 1.02965671 | H |
| −1.98441229 | 0.95231015 | 2.26459774 | H |
| 1.39309678 | −1.99574553 | 0.32548643 | O |
| 2.22341324 | −2.09322248 | 0.80818205 | H |
| 0.89900611 | −2.80678861 | 0.49736120 | H |
| 0.33086908 | 0.32312074 | 0.27861446 | O |
| 0.70909144 | −0.61683840 | 0.32564251 | H |
| 0.30256111 | 0.65973388 | −0.67775409 | H |
| −0.58479628 | 0.37226314 | 0.71210122 | H |

- **2d: $H_3O^+(H_2O)_6$**

|  |  |  |  |
|---:|---:|---:|:---:|
| 1.85150370 | −1.51150331 | 0.90398144 | O |
| 1.24199992 | −2.24293445 | 0.70821760 | H |
| 1.05690465 | 0.97390803 | 0.56629088 | O |
| 1.37430179 | 0.04402253 | 0.71579675 | H |
| −0.00075508 | −3.47576401 | −0.00046705 | O |
| −1.85265994 | −1.51087976 | −0.90289804 | O |
| −1.24306531 | −2.24235603 | −0.70739050 | H |
| −1.05732195 | 0.97365080 | −0.56714678 | O |
| −1.37469838 | 0.04344055 | −0.71601154 | H |
| −0.00126959 | 0.97225982 | −0.00098363 | H |
| −2.29620924 | −1.74939011 | −1.72424813 | H |
| 2.29454890 | −1.74996349 | 1.72561072 | H |
| −1.75379699 | 1.45550637 | −0.05724897 | H |
| 1.75373438 | 1.45541157 | 0.05683420 | H |
| −3.00056516 | 2.26882467 | 0.69832236 | O |
| 3.00181331 | 2.26938348 | −0.69802102 | O |
| 0.43139845 | −4.06735448 | −0.63064479 | H |
| −0.43248901 | −4.06839491 | 0.62902021 | H |
| −3.03395986 | 2.42251534 | 1.64916367 | H |
| −3.34902958 | 3.07577257 | 0.30241737 | H |
| 3.34974969 | 3.07664338 | −0.30205910 | H |
| 3.03700313 | 2.42282383 | −1.64911669 | H |

- **3d: $H_3O^+(H_2O)_6$**

|            |            |            |   |
|-----------:|-----------:|-----------:|---|
| −0.75171871 |  1.03748055 |  2.17496922 | O |
| −0.91681868 |  0.08441846 |  2.03673146 | H |
| −1.49290940 |  1.38324339 |  2.68473933 | H |
|  1.60459680 | −1.61836012 | −0.05789930 | O |
|  2.27633959 | −2.31006042 | −0.01650736 | H |
|  1.03057304 | −1.75450607 |  0.72011123 | H |
| −0.75465541 | −1.61812893 |  1.41667854 | O |
| −1.12733476 | −2.30996368 |  1.97690823 | H |
| −1.14138094 | −1.75296621 |  0.53032758 | H |
|  2.25895811 |  1.03823689 | −0.43457575 | O |
|  2.22199396 |  0.08474490 | −0.22436895 | H |
|  3.07098045 |  1.38338751 | −0.04692310 | H |
| −1.50586904 |  1.04081870 | −1.73716772 | O |
| −1.57604496 |  1.38767029 | −2.63358102 | H |
| −1.30514306 |  0.08751236 | −1.81206991 | H |
| −0.85216773 | −1.61634117 | −1.36368883 | O |
| −1.15237401 | −2.30720330 | −1.96700035 | H |
|  0.10860897 | −1.75277322 | −1.25583773 | H |
|  0.00088154 |  2.10331000 |  0.00205807 | O |
| −0.34773912 |  1.73306953 |  0.88740720 | H |
|  0.94193023 |  1.73330580 | −0.13930929 | H |
| −0.59198861 |  1.73437311 | −0.74281782 | H |

- (OH)$^-$

|            |            |            |   |
|-----------:|-----------:|-----------:|---|
| −0.06682059 | −0.08717002 |  0.00000000 | O |
|  0.52416865 |  0.68379815 |  0.00000000 | H |

- **OH$^-$H$_2$O**

$$\begin{array}{rrrl}
-0.07044820 & -0.13873308 & 0.01705511 & H \\
1.42558909 & 0.55252600 & 0.61705249 & H \\
-1.22470061 & -0.10436082 & 0.06305916 & O \\
1.23345241 & -0.10725632 & -0.06123216 & O \\
-1.39424762 & 0.57584637 & -0.59878353 & H
\end{array}$$

- **OH$^-$(H$_2$O)$_2$**

$$\begin{array}{rrrl}
-1.46840061 & -0.03872027 & 0.01194047 & H \\
-2.45979734 & -0.29357099 & -0.13617348 & O \\
-0.01592555 & 1.30999645 & 0.34712418 & H \\
-2.50799921 & -0.23462969 & -1.09598586 & H \\
-0.00301325 & 0.38316548 & 0.08158792 & O \\
1.47400776 & 0.01404387 & -0.02472025 & H \\
2.47145300 & -0.25849344 & 0.00259523 & O \\
2.50610437 & -0.68777916 & 0.86376835 & H
\end{array}$$

- **OH$^-$(H$_2$O)$_3$**

|            |            |            |   |
|-----------:|-----------:|-----------:|---|
|  0.83685055 | −1.10139974 | −0.97261131 | H |
|  1.61142446 | −0.69338710 | −0.55658142 | O |
|  1.18768028 | −0.45402414 |  0.32181586 | H |
| −0.02554296 | −0.02960602 |  2.32402105 | H |
| −0.20888629 |  1.25824927 |  0.32002112 | H |
|  0.57184245 |  1.26716478 | −0.94499089 | H |
| −0.17843801 |  1.74358297 | −0.55704191 | O |
| −0.01729916 | −0.01849610 |  1.36438083 | O |
| −0.99980285 | −0.81280815 |  0.28898667 | H |
| −1.41294066 | −1.02479862 | −0.60000518 | O |
| −1.36136078 | −0.13634210 | −0.98498241 | H |

- **c4: $OH^-(H_2O)_4$**

|            |            |            |   |
|-----------:|-----------:|-----------:|---|
| −0.10261864 |  1.48145206 |  0.40611453 | H |
|  1.48127349 |  0.10264333 |  0.40631389 | H |

|            |            |            |   |
|-----------:|-----------:|-----------:|---|
|  0.10266276 | −1.48163315 |  0.40645748 | H |
| −1.48115214 | −0.10254151 |  0.40607591 | H |
| −0.00001669 | −0.00004158 |  1.31319492 | O |
| −0.00010231 | −0.00000475 |  2.27375989 | H |
| −0.03742845 |  2.07655322 | −0.38617167 | O |
|  2.07652917 |  0.03771263 | −0.38592147 | O |
|  0.03750440 | −2.07656392 | −0.38624802 | O |
| −2.07666863 | −0.03765563 | −0.38568032 | O |
|  0.73084478 |  1.65891492 | −0.80758252 | H |
|  1.65906554 | −0.73051793 | −0.80757840 | H |
| −0.73080691 | −1.65873041 | −0.80765884 | H |
| −1.65932600 |  0.73037375 | −0.80768221 | H |

- **cs: OH$^-$(H$_2$O)$_4$**

|            |            |            |   |
|-----------:|-----------:|-----------:|---|
|  2.11055955 | −0.01512024 | −0.58090807 | O |
| −0.00023407 | −1.45710756 | −0.55298679 | O |
|  0.76410179 |  1.33175652 | −0.48873064 | H |
|  2.20505301 | −0.00119905 |  0.38052329 | H |
|  0.00024856 |  1.88644428 | −0.23033206 | O |
| −0.76340752 |  1.33200728 | −0.48980332 | H |
| −2.11053892 | −0.01445543 | −0.58385004 | O |
| −2.20662936 | −0.00055493 |  0.37734307 | H |
| −1.33465083 | −0.67020586 | −0.67785565 | H |
| −0.00113045 | −0.25655418 |  1.84547483 | O |
| −0.00087273 |  0.61241882 |  1.40997629 | H |
| −0.00094576 | −0.85721320 |  1.05209234 | H |
|  1.33458916 | −0.67059398 | −0.67609181 | H |
| −0.00020409 | −2.40388383 | −0.71419341 | H |

- **OH$^-$(H$_2$O)$_5$**

|           |           |           |   |
|----------:|----------:|----------:|---|
|  1.49631300 | −0.42419396 |  0.00959014 | H |
|  0.07630620 | −1.95313021 | −0.03980819 | H |
| −1.19703327 | −0.40301454 |  0.46859894 | H |
|  0.08186169 |  1.84767393 | −1.08989347 | O |
|  1.87493142 | −0.30350332 | −0.90608108 | O |
| −0.19199518 | −2.37128472 | −0.89693387 | O |
| −1.87760824 | −0.06687056 | −0.16153370 | O |
|  0.76189596 |  1.14248576 | −1.15359847 | H |
|  1.37894666 | −1.00284146 | −1.36222487 | H |
| −0.96606969 | −1.81966423 | −1.09212927 | H |
| −1.35381492 |  0.59233511 | −0.65693910 | H |
|  0.37370682 | −0.73554524 |  1.21579179 | O |
|  0.57479759 | −1.20414227 |  2.03087111 | H |
| −0.24711347 |  1.76066308 |  1.74636286 | O |
|  0.06791190 |  2.01999223 | −0.12791598 | H |
| −1.19090663 |  1.65334374 |  1.56776274 | H |
|  0.07077458 |  0.81067795 |  1.62359300 | H |

- **OH$^-$(H$_2$O)$_6$**

|  |  |  |  |
|---:|---:|---:|:---:|
| 0.09528687 | −1.70714985 | −1.72030904 | O |
| 0.35048692 | −1.13937640 | −2.45841762 | H |
| −0.61802117 | −1.14956772 | −1.27198454 | H |
| −0.71146925 | −1.87504736 | 1.34966271 | O |
| −0.12553300 | −2.36756114 | 0.76133104 | H |
| −1.14503277 | −1.23175591 | 0.70903305 | H |
| −0.92151680 | 2.04054276 | 0.90337797 | O |
| −1.31182698 | 1.21957397 | 0.49471558 | H |
| −0.39382703 | 2.34944383 | 0.14945771 | H |
| 0.91047190 | 0.29939329 | 2.11329073 | O |
| 0.36817547 | −0.50819802 | 1.99966175 | H |
| 0.30348660 | 1.00070236 | 1.79208193 | H |
| 0.21218505 | 1.49326714 | −1.76345053 | O |
| 0.98337379 | 1.07849471 | −1.32847277 | H |
| −0.52497758 | 0.94200819 | −1.41492823 | H |
| 2.10090511 | −0.14639085 | −0.37128634 | O |
| 1.76477751 | 0.03021539 | 0.53554513 | H |
| 1.48780035 | −0.82563323 | −0.70976181 | H |
| −1.56717100 | −0.16383071 | −0.43767017 | O |
| −2.48146301 | −0.26874742 | −0.71682925 | H |

# TOC Graphic

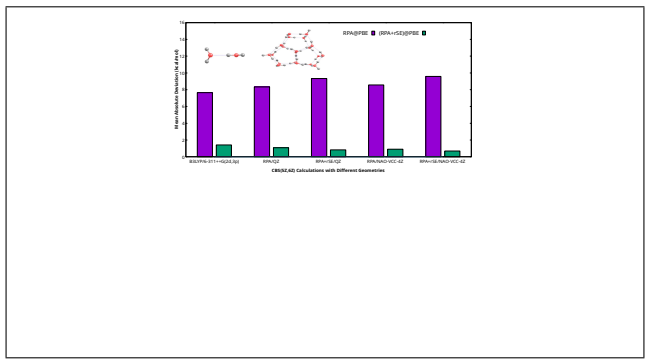

	}

\end{document}